\documentclass[11pt,x11names,a4paper]{article}

\usepackage[T1]{fontenc} 
\usepackage{lmodern}
\usepackage{microtype} 

\usepackage[a4paper, left=25mm, right=25mm, top=30mm, bottom=25mm]{geometry} 
\usepackage{fancyhdr} 
\usepackage{abstract}

\usepackage{titlesec} 
\titleformat{\section}[block]{\large\bfseries\centering}{\thesection}{1em}{} 
\titleformat{\subsection}[block]{\bfseries}{\thesubsection}{1em}{} 

\usepackage{xcolor}

\usepackage{cite}
\usepackage{hyperref}
\hypersetup{
	colorlinks=true,
	linkcolor=Blue4,
	citecolor=Red4,
	urlcolor=Green4,
	linktoc=page
}

\usepackage{enumerate}
\usepackage{graphicx}
\usepackage[percent]{overpic}

\usepackage{booktabs}
\usepackage{bbm}
\usepackage[margin=15pt,small]{caption}

\usepackage{amsmath,amssymb,slashed}
\usepackage{amsthm}

\numberwithin{equation}{section}
\usepackage{slashed}

\usepackage{subcaption} 

\newcommand{\bb}[1]{\mathbb{#1}}
\newcommand{\dd}{\mathrm{d}}

\newcommand{\e}{\mathrm{e}}
\newcommand{\rmi}{\mathrm{i}}
\newcommand{\rme}{\mathrm{e}}
\newcommand{\rmd}{\mathrm{d}}

\newcommand{\be}{\begin{equation}}
	\newcommand{\ee}{\end{equation}}
\newcommand{\bea}{\begin{eqnarray*}}
	\newcommand{\eea}{\end{eqnarray*}}

\newcommand{\f}[2]{\frac{#1}{#2}}
\newcommand{\R}{\mathbf{R}}
\newcommand{\Z}{\mathbf{Z}}

\newcommand{\vol}{\text{vol}}
\renewcommand{\Re}{\text{Re}}
\renewcommand{\Im}{\text{Im}}

\newcommand{\Tr}{\text{Tr}~}

\newcommand{\SU}{\mathop{\rm SU}}
\newcommand{\SO}{\mathop{\rm SO}}
\renewcommand{\O}{\mathop{\rm O}}
\newcommand{\SL}{\mathop{\rm SL}}
\newcommand{\U}{\mathop{\rm {}U}}

\newcommand{\ssl}{\mathfrak{sl}}
\newcommand{\Ess}{\text{E}_{7(7)}}

\newcommand{\ess}{\mathfrak{e}_{7(7)}}

\title{\fontsize{20pt}{24pt}\selectfont\textbf{The Conformal Manifold of S-folds \\in String Theory}\vspace{2mm}}

\author{
\large{
	\href{mailto:nikolay.bobev@kuleuven.be}{Nikolay Bobev},$^1$ \href{mailto:ffg@hi.is}{Fri{\dh}rik Freyr Gautason},$^2$ and \href{mailto:jvanmuid@sissa.it}{Jesse van Muiden}$^{3,4}$}\\[5mm]
{}$^1${\normalsize KU Leuven, Instituut voor Theoretische Fysica}\\
{\normalsize Celestijnenlaan 200D, B-3001 Leuven, Belgium}\\[5mm]
{}$^2${\normalsize University of Iceland, Science Institute}\\
{\normalsize Dunhaga 3, 107 Reykjav{\'i}k, Iceland}\\[5mm]
{}$^3${\normalsize SISSA}\\
{\normalsize Via Bonomea 265, I 34136 Trieste, Italy}\\[5mm]
{}$^4${\normalsize INFN - Sezione di Trieste}\\
{\normalsize Via Valerio 2, I-34127 Trieste, Italy}
}

\date{}

\begin{document}
{\hypersetup{urlcolor=black}\maketitle}
\thispagestyle{empty}
\begin{abstract}
\noindent We continue the holographic exploration of the conformal manifold of 3d $\mathcal{N}=2$ S-fold SCFTs constructed by gauging the flavor symmetry of the Gaiotto-Witten $T[U(N)]$ theory.  We show how to uplift the two-parameter family of AdS$_4$ vacua dual to this conformal manifold to 10d backgrounds of type IIB supergravity. We use these uplifted solutions to shed new light on the mysterious nature of the infinite distance limit on the conformal manifold and to study probe strings and D3-branes. This analysis uncovers an intriguing structure of the $S^3$ partition function of the S-fold SCFTs which resembles the giant graviton expansion of the superconformal index of 4d $\mathcal{N}=4$ SYM. We also show how to each member of the family of supersymmetric AdS$_4$ vacua one can associate a consistent truncation to 4d $\mathcal{N}=2$ gauged supergravity and use this result, in conjunction with holography, to calculate the large $N$ partition function of the 3d S-fold SCFT on compact Euclidean manifolds. Finally, we generalize the supersymmetric AdS$_4$ vacua to a four-parameter family of non-supersymmetric AdS$_4$ solutions.

\end{abstract}
\newpage
\tableofcontents
\section{Introduction and summary}

In this work we explore an interesting example of the AdS$_4$/CFT$_3$ correspondence in type IIB supergravity that arises from $N$ D3-branes on $S^1$ with an $\SL(2,\mathbf Z)$ S-duality twist along the circle. The backreaction of the branes produces an AdS$_4$ solution of type IIB supergravity that preserves $\mathcal{N}=4$ supersymmetry and has an internal space of the schematic form $S^1 \times S^5$ with an $\SL(2,\mathbf Z)$ monodromy of the axio-dilaton along the $S^1$ \cite{Inverso:2016eet,Assel:2018vtq}. The 3d CFT holographically dual to this AdS$_4$ vacuum can be obtained by starting with the $T[U(N)]$ theory of Gaiotto-Witten \cite{Gaiotto:2008ak,Gaiotto:2008sd} and gauging its $\U(N)\times \U(N)$ flavor symmetry with a vector multiplet and adding a Chern-Simons term at level $k$ for the gauge field together with a specific superpotential \cite{Assel:2018vtq}. Despite the non-geometric features of the string theory background and the strongly coupled nature of the dual SCFT, this model can be studied quantitatively with various techniques and admits interesting generalizations and extensions. 

The $\mathcal{N}=4$ AdS$_4$ solution of \cite{Inverso:2016eet,Assel:2018vtq} belongs to a two-dimensional family of supersymmetric AdS$_4$ vacua of type IIB supergravity that preserve $\mathcal{N}=2$ supersymmetry. These solutions can be constructed using the 4d $\mathcal{N}=8$ gauged supergravity with $[\SO(6)\times \SO(1,1)] \ltimes \mathbf R^{12}$ gauging, see \cite{Guarino:2020gfe,Bobev:2021yya}. Alternatively, one can use the 5d $\mathcal{N}=8$ $\SO(6)$ gauged supergravity and construct these supersymmetric AdS$_4$ solutions as limits of gravitational solutions describing Janus interfaces \cite{Bobev:2020fon,Arav:2021gra}. Both of these perspectives and the consistent truncation results for the 4d and 5d gauged supergravity theories in \cite{Inverso:2016eet} and \cite{Baguet:2015sma}, respectively, guarantee that the AdS$_4$ vacua can be uplifted to full-fledged backgrounds of type IIB string theory. Performing these uplifts explicitly is an arduous exercise which we perform in detail below and obtain the full 10d uplift of the two-parameter family of AdS$_4$ $\mathcal{N}=2$ supersymmetric vacua of \cite{Bobev:2021yya}. In addition to serving as rare explicit examples of non-geometric backgrounds in string theory, this family of solutions is of interest in the context of the AdS/CFT correspondence.

The ten-dimensional type IIB backgrounds we construct should be thought of as arising from a stack of $N$ D3-branes wrapped on $S^1$ with an $\SL(2,\mathbf Z)$ monodromy along the circle. The background takes the schematic form
\begin{equation}\label{eq:10dbackgrintro}
	w\text{AdS}_4 \times S^1_{\mathfrak J} \times \tilde S^5\,,
\end{equation}
where $\mathfrak J$ denotes the $\SL(2,\mathbf Z)$ element, $\tilde S^5$ is a squashed five sphere, and there is warp factor in front of the AdS$_4$ part of the metric that depends on the coordinates on the sphere. We note that this type IIB background is non-geometric since $S^1_{\mathfrak J}$ is made compact by using the $\SL(2,\mathbf Z)$ symmetry of type IIB string theory. The ten-dimensional background has all NS-NS and R-R fluxes turned on and the squashed metric on $\tilde S^5$ preserves only $\U(1) \times \U(1)$ isometry. This isometry corresponds to the $\U(1)_R$ R-symmetry and $\U(1)_F$ flavor symmetry preserved along the conformal manifold in the dual 3d $\mathcal{N}=2$ SCFT. The number of D3-branes, as determined by the flux quantization of the five-form flux, and the periodicity of the circle $S^1_\mathfrak{J}$ determine the rank $N$ of the gauge group and the corresponding CS level $k$ in the SCFT, respectively. There are two special members of the family of supersymmetric AdS$_4$ backgrounds that enjoy enhanced symmetry. One of these solutions has $\SU(2)\times \SU(2)$ isometry on $\tilde S^5$ and $\mathcal N=4$ supersymmetry and corresponds to the background discussed in \cite{Inverso:2016eet,Assel:2018vtq}, while the other has $\SU(2)\times \U(1)$ isometry, $\mathcal N=2$ supersymmetry, and is the same as the AdS$_4$ vacuum presented in \cite{Guarino:2020gfe,Bobev:2020fon}.

These features of the family of ten-dimensional backgrounds mirror the properties of the conformal manifold in the dual S-fold SCFT. A convenient way to think about this conformal manifold is to remember that upon gauging the $\U(N)\times \U(N)$ flavor symmetry symmetry of the $T[U(N)]$ $\mathcal{N}=4$ SCFT the conformal symmetry is broken and an RG flow ensues. As argued in \cite{Assel:2018vtq} the IR theory is conformal and furthermore there is a freedom in the choice of IR superpotential that can be expressed schematically as, see \cite{Bobev:2021yya},
\begin{equation}\label{eq:WIRintro}
	W_{\text{IR}} = -\frac{2\pi}{k} \Tr (\mu_H \mu_C) + \lambda \,\Tr (\mu_H\mu_C)\,,
\end{equation}
where $\mu_H$ and $\mu_C$ are the two moment map operators associated with the $\U(N)\times \U(N)$ flavor symmetry of the $T[U(N)]$ theory. Here $k$ is the CS level of the gauge group and $\lambda$ is a complex parameter that acts as an exactly marginal coupling preserving $\mathcal{N}=2$ supersymmetry. For generic values of $\lambda$ the symmetry of the $\mathcal{N}=2$ SCFT is $\U(1)_R \times \U(1)_F$. At $\lambda=0$ one finds enhancement to $\mathcal{N}=4$ superconformal symmetry and its associated $\SU(2)\times \SU(2)$ R-symmetry. Despite the strongly-coupled nature of this family of SCFTs, it is possible to use supersymmetric localization to calculate the $S^3$ partition function of the theory. The result is remarkably simple and reads, see \cite{Assel:2018vtq,Ganor:2014pha,Terashima:2011qi},
\begin{equation}\label{Eq: susy partition function}
	\mathcal Z_N = \frac{q^{N^2/2}}{(q)_N}\,,
\end{equation}
where $q = \rme^{-T}$, $k = 2\cosh T$, and $(q)_N$ is the $q$-Pochhammer symbol, see Section~\ref{Sec: non perturbative contributions} below for further details. As emphasized in \cite{Bobev:2021yya} this partition function is independent of the marginal coupling $\lambda$ and is therefore the same over the whole conformal manifold. Given this, and the scarcity of other QFT techniques to study the properties of conformal manifolds in 3d $\mathcal{N}=2$ SCFTs, one can resort to the supergravity description to find further insights into the properties of this model.

Using the 4d maximal gauged supergravity construction of the family of $\mathcal{N}=2$ AdS$_4$ vacua it was shown in \cite{Bobev:2021yya} that the Zamolodchikov metric of the conformal manifold takes the form 
\begin{equation}\label{Eq: Zamolodchikov metric}
	\dd s_Z^2 = \frac{4-Y^2}{2}\left(\frac{\dd Y^2}{Y^2(2-Y^2)} + d\chi^2\right)\,.
\end{equation}
Here $Y$ and $\chi$ are conveniently chosen combinations of the massless 4d supergravity scalar fields and this result should be viewed as a supergravity prediction for the large $N$ limit of the Zamolodchikov metric. The values of $Y$ are constrained by the structure of 4d supergravity to lie in the interval $[0,\sqrt{2}]$, while a priori the supergravity scalar $\chi$ can take any real value. A highly non-trivial calculation of the type IIB supergravity KK spectrum as a function of $\chi$ can be performed using techniques from Exceptional Field Theory \cite{Malek:2019eaz}. The details of this analysis reveal that $\chi$ should be periodically identified and should lie in the range $[0,2\pi/T)$ with $T$ defined below \eqref{Eq: susy partition function}, see \cite{Giambrone:2021zvp,Guarino:2021kyp}. A similar KK spectroscopy analysis can be performed along the entire two-dimensional space of vacua parametrized by $Y$ and $\chi$ but it does not reveal any structure that may suggest additional periodic identifications \cite{Bobev:2021yya,Cesaro:2021tna}. We note that in this parametrization the $\mathcal{N}=4$ AdS$_4$ vacuum is at $Y=1$ and $\chi=0$, while the $\mathcal{N}=2$ AdS$_4$ solution with enhanced $\SU(2)\times U(1)$ symmetry is at $Y=\sqrt{2}$ and $\chi=0$.

The supergravity analysis above therefore suggests that at large $N$, the conformal manifold has the topology of a cylinder and moreover one can show that the metric in \eqref{Eq: Zamolodchikov metric} features unbounded geodesics. This may lead to the puzzling conclusion that the conformal manifold is non-compact. As discussed in \cite{Perlmutter:2020buo,Baume:2023msm}, non-compact conformal manifolds are usually equipped with an infinite distance locus where the CFT at hand becomes free. The analog of such a locus in the S-fold conformal manifold discussed above is at $Y \to 0$. Indeed, this appears to be a singular limit in the 5d gauged supergravity construction of the family of AdS$_4$ solutions. However, the KK spectrum of supergravity excitations does not reveal the emergence of higher spin conserved currents associated with a free sector in the CFT. This is furthermore supported at finite $N$ by the $S^3$ partition function in \eqref{Eq: susy partition function} which does not have the structure of the partition function of free 3d $\mathcal{N}=2$ SCFTs. We are thus led to the conclusion that either the conformal manifold is non-compact with unbounded geodesics but without a free locus, or there is a tower of light higher-spin modes that arises in the $Y \to 0$ locus and remains inaccessible in supergravity. The 10d uplift of the family of AdS$_4$ vacua that we present in this work sheds new light on this mysterious feature of the conformal manifold. We find that the singular behavior of the 5d gauged supergravity in the limit $Y \to 0$ becomes much more mild in the 10d background. As $Y \to 0$ the $w$AdS$_4$ and $\tilde S^5$ parts of the 10d metric, see \eqref{eq:10dbackgrintro}, remain regular and finite size, while the radius of the non-geometric circle $S^1_{\mathfrak J}$ blows up as $1 /Y^2$. We do not know how to resolve this singularity and have not completed the study of the spectrum of light string or D-brane modes in the $Y\to0$ limit and so the question whether this conformal manifold is compact or whether it has a free locus remains open. Nevertheless, we cannot resist to make a few speculations and observations. It appears that in the $Y \to 0$ limit the period of the S-fold circle blows up. Taking this as a starting point, we note that since the periodicity of the circle is related to the CS level $k$ (as described above), one can conclude that this limit can be offset by an appropriate change of the CS level $k$. Perhaps this points to a dual description of the SCFT in terms of a different theory with a large CS level but we were unable to come up with a concrete proposal for such a theory. It will be very interesting to explore this question further and understand better the global properties of this conformal manifold of S-fold SCFTs. 

While the main focus of our work is on the two-parameter family of $\mathcal{N}=2$ supersymmetric AdS$_4$ vacua described above, as a byproduct of our calculations we also present a new four-parameter family of non-supersymmetric AdS$_4$ solutions. As discussed in \cite{Guarino:2019oct,Bobev:2019jbi,Guarino:2020gfe,Bobev:2020fon,Arav:2021tpk,Guarino:2021hrc} one can use 4d and 5d maximal gauged supergravity to construct S-fold AdS$_4$ solutions that preserve only $\mathcal{N}=1$ supersymmetry or break it completely. Furthermore, it was shown in \cite{Guarino:2021hrc} that for S-fold AdS$_4$ solutions with continuous global symmetry, one can introduce new parameters that act as a marginal deformations that potentially break supersymmetry. We show below how to use 4d maximal gauged supergravity in combination with the method of \cite{Guarino:2021hrc} to construct a four-parameter family of such AdS$_4$ solutions. For general values of the parameters, the solutions break supersymmetry completely but on a two-dimensional subspace the solutions reduce to the family of ${\cal N} = 2$ AdS$_4$ S-folds described above. One could reasonably wonder whether the non-supersymmetric AdS$_4$ solutions we construct are stable. A setup similar to ours was studied in \cite{Giambrone:2021wsm}, where it was argued that a similar family of non-supersymmetric AdS$_4$ S-fold solutions is stable against many perturbative and non-perturbative potential decay channels. Much of the reasoning \cite{Giambrone:2021wsm} relies on the fact that the non-supersymmetric vacua are continuously connected through 10d diffeomorphisms to the supersymmetric vacua. This feature is also exhibited by the four-parameter family of solutions we construct here and thus we could import the conclusions of \cite{Giambrone:2021wsm}. It will of course be very interesting to study in more detail the perturbative and non-perturbative stability of our solutions but we leave this important question to future work.

The $S^3$ partition function \eqref{Eq: susy partition function} of the S-fold SCFTs we study has a simple form that reveals an interesting structure. We take inspiration from the well-known relation between the 4d $\mathcal{N}=1$ superconformal index and the 3d $\mathcal{N}=2$ $S^3$ partition function for certain classes of supersymmetric QFTs discussed in \cite{Dolan:2011rp,Gadde:2011ia,Imamura:2011uw,Aharony:2013dha} to think of the result in \eqref{Eq: susy partition function} as a limit of some type of a 4d partition function. Given the D3-brane origin of the S-fold SCFTs we discuss, it is natural to conjecture that this is the partition function of the 4d $\mathcal{N}=4$ SYM theory on $S^1_{\mathfrak J} \times S^3$ where one performs an S-duality twist, specified by the $\SL(2,\mathbf Z)$ element ${\mathfrak J}$, of the 4d theory as one goes along the circle. To the best of our knowledge such S-duality ``twisted indices'' have not been studied in the literature and it will be very interesting to understand whether they can be properly defined and calculated. Undeterred by the lack of a proper QFT treatment of these putative new indices we show that indeed \eqref{Eq: susy partition function} can be rewritten in a way reminiscent of the 1/2-BPS superconformal index of the 4d $\mathcal{N}=4$ SYM. We do this by exploiting the so-called giant graviton expansion of the 4d $\mathcal{N}=4$ SYM index which is tailored to holography and organizes the index as a series of contributions from BPS operators associated to D3-branes, see \cite{Biswas:2006tj,Arai:2019xmp,Imamura:2021ytr,Gaiotto:2021xce}. We show how the $S^3$ partition function of the S-fold theories in \eqref{Eq: susy partition function} can be written in a similar fashion. Moreover, we study probe supersymmetric D3-branes in the 10d type IIB supergravity solutions we construct and show that their on-shell action agrees with the structure of the ``giant graviton expansion'' of the $S^3$ partition function. This result is a non-trivial test of the holographic correspondence for the S-fold SCFT and their interpretation in terms of D3-branes. Moreover, it underscores the possible relation between the S-fold $S^3$ partition function and a new S-duality ``twisted index'' of the 4d $\mathcal{N}=4$ SYM theory.

As an additional check of the holographic duality in our setup we also explore supersymmetric line operators in the S-fold SCFTs. We first study probe $(p,q)$-strings with an AdS$_2$ worldvolume in the family of AdS$_4$ type IIB solutions we construct. We compute their regularized on-shell action and show that it is independent of the parameters $Y$ and $\chi$. This implies that these probe strings should be dual to supersymmetric line operators in the S-fold theory on $S^3$. The vacuum expectation values of such line operators have not been studied in the literature, although it should be possible to do so using supersymmetric localization. We eschew a detailed supersymmetric localization analysis in favor of a simple Gaussian matrix model argument with which we are able to estimate the expectation value of a supersymmetric Wilson loop along the equator of $S^3$ in the fundamental representation of the gauge group. Despite this crude matrix model approximation we find very good agreement with the probe string analysis on the gravity side.

It is natural to wonder whether other supersymmetric observables in the S-fold SCFTs can be computed using supersymmetric localization. While there are some results on the 3d superconformal index available in the literature, see \cite{Bachas:2019jaa,Beratto:2020qyk,Garozzo:2019ejm}, this question is largely unexplored due to the difficulties presented by the non-Lagrangian nature of the $T[U(N)]$ theory. In the large $N$ limit, supergravity can again be brought to bear on this problem and leads to interesting results. We show that for every allowed value of the parameters $Y$ and $\chi$ one can find a consistent truncation of type IIB supergravity to the 4d $\mathcal{N}=2$ minimal gauged supergravity which has a bosonic sector consisting of the metric and a $\U(1)$ Maxwell field. This simple supergravity theory is expected to describe holographically the universal large $N$ dynamics of the energy-momentum multiplet in 3d  holographic $\mathcal{N}=2$ SCFT. This result can be used to compute the large $N$ behavior of supersymmetric partition functions of the S-fold SCFTs on a variety of Euclidean 3-manifolds $\mathcal{M}_3$. The calculation leverages available results in the literature for the regularized on-shell action of Euclidean gauged supergravity solutions that can be viewed as smooth fillings of $\mathcal{M}_3$. Using this approach we present explicit results for the large $N$ squashed sphere partition function, topologically twisted index, and superconformal index of the S-fold SCFTs. As expected from supersymmetric localization, our results do not depend on the exactly marginal couplings and are thus valid along the entire conformal manifold. Our analysis also reveals the existence of asymptotically AdS$_4$ black hole solutions in the S-fold compactifications of type IIB supergravity and provide simple expressions for their Bekenstein-Hawking entropy.  

We continue in the next section with a summary of the gauged supergravity description of the family of AdS$_4$ vacua of interest as well as its explicit uplift to 10d solutions of type IIB supergravity. In Section~\ref{Sec: non perturbative contributions} we show that the $S^3$ partition function of the S-fold SCFT can be written in a suggestive way resembling the giant graviton expansion familiar from 4d $\mathcal{N}=4$ SYM and support this interpretation with a probe D3-brane calculation. In Section~\ref{sec:hololines} we study probe strings in the type IIB supergravity background and discuss their dual SCFT interpretation as supersymmetric line operators. We discuss how to construct a consistent 4d minimal gauged supergravity truncation associated to each of the AdS$_4$ vacua we study in Section~\ref{sec:S-holes} and explore the consequences of this result for holography. In the two appendices we provide more details on the construction of the supersymmetric family of AdS$_4$ vacua, its non-supersymmetric generalization, as well as the infinite distance limit of the uplifted 10d supergravity solutions. Along with this paper we include an ancillary Mathematica file that contains some details on the 10d type IIB supergravity backgrounds we construct.

\section{S-fold conformal manifold in type IIB}\label{Sec: The novel type IIB S-fold backgrounds}
To construct the IIB supergravity solutions of interest we start from the five-dimensional realization of the one parameter family presented in \cite{Arav:2021gra} and apply the uplift formulae in \cite{Baguet:2015sma}. In ten-dimensions it is then straight-forward to add two additional parameters as locally they can be expressed as coordinate transformations \cite{Giambrone:2021zvp}. This will result in a three-parameter family of 10d solutions of which there is a two-parameter sub-family of type IIB AdS$_4$ solutions that preserve ${\cal N}=2$ supersymmetry. This two-parameter family of supersymmetric solutions will form the backbone of the analysis in the rest of the paper. An alternative way to construct this family of supersymmetric solutions would be to directly uplift the two-parameter family of supersymmetric solutions in four-dimensional supergravity found in \cite{Bobev:2021yya} using the uplift formulae in \cite{Inverso:2016eet}. Presumably this alternative method will lead to the same results as the ones we present below.

\subsection{Review of the five-dimensional solution}\label{Sec: 5d}
The five-dimensional realization of a one-parameter family of AdS$_4$ solutions was constructed in \cite{Arav:2021gra} by starting with a ten-scalar subsector of the maximal $\SO(6)$ gauged supergravity theory in five dimensions.
This $\mathcal{N}=8$ theory has  $42$ scalar fields which often presents difficulties when trying to construct explicit solutions. For this reason one usually employs a convenient consistent truncation to a subset of the scalar fields. In \cite{Bobev:2016nua}, see also \cite{Bobev:2020ttg}, such a ten-scalar consistent truncation was found by keeping only 5d supergravity fields that are invariant under a  $\mathbf Z^3_2$  subgroup of the $\SO(6) \times \SL(2,\mathbf R)$ symmetry of the maximal supergravity theory. The bosonic field content of this truncated theory consists of the five-dimensional metric, together with two real scalars; $\beta_1,\beta_2$ and four complex scalar field $z^i$ with $i=1,\dots,4$. In \cite{Arav:2021gra} a further six-scalar truncation of this model was considered where 
\be\label{sixscalartrunc}
z^2=\bar{z}^2 = -z^4=-\bar{z}^4\,,\qquad \beta_2 =0\,.
\ee
Writing the remaining scalar fields in terms of real variables
\be
\begin{split}
z^1&=\tanh\Big[ \f12( 2\alpha_1+\alpha_3+\varphi-2i\phi_1 +2i\phi_4 ) \Big]\,,\\
z^2&=\tanh\Big[ \f12( \alpha_3-\varphi) \Big]\,,\\
z^3&=\tanh\Big[ \f12( 2\alpha_1-\alpha_3+\varphi-2i\phi_1 -2i\phi_4 ) \Big]\,,
\end{split}
\ee
the one-parameter family of AdS$_4$ vacua of interest here can be found by imposing $\alpha_1=\alpha_3=0$ and
\be\label{5Dscalarsol}
\e^{-12\beta_1} = 2\tan^22\phi_4\equiv Y^2\,,\qquad \cos4\phi_4+2\cos4\phi_1=1\,.
\ee
Furthermore the dilaton is linear along the fifth direction of space-time and can therefore, by a simple coordinate transformation, be used as a convenient coordinate itself. Explicitly the metric takes the form 
	\begin{equation}\label{Eq: five-dimensional metric}
		\rmd s_{5}^2 = L^2 \Big( \frac{Y^{2/3}}{2} \dd s_{\text{AdS}_4} +\frac{1}{Y^{4/3}} \dd \varphi^2 \Big)\,,
	\end{equation}
where we have slightly changed notation compared to \cite{Arav:2021gra}\footnote{In \cite{Bobev:2021yya} these solutions were studied from a four-dimensional perspective. The relation of $Y$ to the parameter used there is $\varphi_{\text{\hspace{-0.1pt}\cite{Bobev:2021yya}}} = \sqrt{2-Y^2}/Y$. This parameterization should not be confused with the dilaton in this paper.} and in particular we work with mostly plus signature for the metric. Note that here $L=2/g$ where $L$ is the length scale of the maximally symmetric AdS$_5$ vacuum solution and $g$ is the five-dimensional gauge coupling.
	
The solution preserves ${\cal N}=2$ supersymmetry and should therefore be holographically dual to a family of 3d $\mathcal{N}=2$ SCFT with its associated $\U(1)_R$ symmetry. In addition to this R-symmetry the family of AdS$_4$ vacua exhibits an additional $\U(1)_F$ flavor symmetry. The parameter $Y$ is unfixed and does not show up in the supergravity potential. This parameter is holographically dual to an exactly marginal operator in the field theory. Regularity of the five-dimensional metric and scalars implies that we must constrain $Y$ to lay in the range $[0,\sqrt{2}]$.

A priori, the dilaton and the coordinate direction it parametrizes seem to be non-compact. When the dilaton grows without bound however this would ultimately lead to an inconsistent supergravity solution since the string coupling blows up. We must therefore  impose that the dilaton remains finite and that $\varphi$ describes a compact direction. To do this we impose that the coordinate direction parametrized by the dilaton is compact and simultaneously apply an $\SL(2,\mathbf Z)$ action to glue the dilaton together resulting in an $S$-fold string theory background. The $\SL(2,\mathbf Z)$ transformation acts on the five-dimensional scalars coset as, see \cite{Bobev:2020fon},
	\begin{equation}
		\mathcal V(\varphi + T) = \mathcal V(\varphi) \mathfrak J\,, \quad \text{where} \quad \mathfrak J = \left(\begin{array}{cc}
			\rme^{-T} & 0 \\
			0 & \rme^{T}
		\end{array}\right)\,,
	\end{equation}
and $T$ is the period of the $S^1_\mathfrak J$ circle parametrized by $\varphi$. As written $\mathfrak J$ is an element of $\SL(2,\mathbf R)$ and not $\SL(2,\mathbf Z)$. This can be remedied by employing a  similarity transformation such that the element $\mathfrak J \in \SL(2,\mathbf Z)$ is
	\begin{equation}\label{eq:Jkmatrixdef}
		\mathfrak J_k = \left(\begin{array}{cc}
			k & 1 \\
			-1 & 0
		\end{array}\right)\,,
	\end{equation}
	where $k = 2 \cosh T\in {\bf Z}$. As discussed in the introduction, this integer is identified with a CS level in the dual three-dimensional field theory. Regularity of the background imposes that $k$ is bounded from below $k>2$.

In \cite{Arav:2021gra,Bobev:2021yya} the on-shell action of this family of AdS$_4$ solutions was shown to be independent of $Y$ consistent with the fact that this parameter is dual to an exactly marginal deformation. The value of the on-shell action is\footnote{This  holds for the full two-dimensional family of vacua studied in \cite{Bobev:2021yya}.}
	\begin{equation}\label{eq:SonshellAdS4}
		S_\text{on-shell} = \frac{N^2 T}{2}\,,
	\end{equation}
	which matches the leading order term in the large $N$ limit of the field theory supersymmetric partition function in equation \eqref{Eq: susy partition function}.

At two special values of the parameter $Y$ there is enhanced symmetry. At $Y=1$  the solutions preserves $\mathcal{N}=4$ supersymmetry and the global symmetry is enhanced to $\SO(4)$, this is the original ${\cal N}=4$ S-fold background found in \cite{Inverso:2016eet,Assel:2018vtq}. The second point of interest is $Y=\sqrt{2}$ where the $\U(1)_F$ flavor symmetry is enhanced to $\SU(2)$. This flavor symmetry enhanced point plays an important role in our discussions since its ten-dimensional uplifted solution was already constructed in \cite{Bobev:2020fon}. The new type IIB solution discussed below share many features with the solution in \cite{Bobev:2020fon} and will act as an important consistency check of our calculations. 

The limit $Y\rightarrow 0 $ corresponds to taking an infinite distance limit on the space of massless supergravity scalars which in turn corresponds to an infinite distance limit on the conformal manifold of the dual 3d $\mathcal{N}=2$ SCFT. This point is of special interest in this work and will be discussed further below. Here we note that the 5d metric in \eqref{Eq: five-dimensional metric} becomes singular as $Y\rightarrow 0 $. One of the results we present below is that this singular behavior becomes more mild in the 10d uplift of this family of solutions dimensions.

\subsection{One parameter family in type IIB}\label{Sec: One parameter family of novel type IIB vacua}

To uplift the five-dimensional family of AdS$_4 \times  S^1_{\mathfrak J}$ vacua we use the techniques of exceptional field theory described in \cite{Baguet:2015sma}. The only input needed for this calculation is the 5d solution in \eqref{sixscalartrunc}, \eqref{5Dscalarsol}, and \eqref{Eq: five-dimensional metric}.\footnote{We follow the conventions described in \cite{Bobev:2018hbq} for the explicit uplift.} As discussed in the beginning of this section, the family presented in Section~\ref{Sec: 5d} is a one-dimensional subset of the full two-dimensional family of solutions first found in \cite{Bobev:2021yya}. First we uplift the one-parameter family controlled by $Y$, which will already prove to be formidable, and then in Section~\ref{Sec: Two parameter family of novel type IIB vacua} we discuss how to include two additional parameters based on symmetry considerations. We will see that for generic values of the three parameters, all supersymmetries are broken. However, by setting  one of the three parameters to zero, we find the ten-dimensional uplift of the set of supersymmetric solutions described in \cite{Bobev:2021yya}.\footnote{In Appendix~\ref{App: non-susy sols} we provide a four dimensional extension of the solution in \cite{Bobev:2021yya} using four-dimensional gauged supergravity. We do not uplift this extended solution to ten-dimensions.}

Because the solutions of interest generically only preserve a $\U(1)\times \U(1)$ isometry in the internal space, the expressions for the 10d fields will be rather involved. Nevertheless, we are able to write the full solution analytically as a function of $Y$. For convenience we have included a Mathematica notebook with the full type IIB solution as an ancillary file along with the arXiv submission.

We start with the embedding coordinates used to describe the five-sphere
\begin{equation}
\begin{aligned}
	&\mathcal Y^1 + \rmi \mathcal Y^3 = \rme^{-\rmi (\xi_1 + \xi_2)/2} \cos\theta \cos\xi_3 \,,\\
	&\mathcal Y^2 + \rmi \mathcal Y^4 = \rmi \rme^{-\rmi (\xi_1-\xi_2)/2} \cos\theta\sin\xi_3 \,,\\
	&\mathcal Y^5 +\rmi \mathcal Y^6 = \rme^{-\rmi \phi} \sin\theta\,.		
\end{aligned}
\end{equation}
In these coordinates the metric on the round five-sphere is given by
\begin{equation}
	\rmd s_{S^5}^2 = \rmd \theta^2 + \sin^2\theta \rmd \phi^2 + \cos^2\theta ( \sigma_1^2 + \sigma_2^2 + \sigma_3^2 )\,,
\end{equation}
where
\begin{equation}\label{Eq: su2 one forms}
	\sigma_1+\rmi \sigma_2 = -\rme^{-\rmi \xi_2} (\rmi \rmd \xi_3 - \frac12 \sin2\xi_3 \rmd \xi_1)\,,\quad \sigma_3 = -\frac12 (\rmd \xi_2 + \cos 2\xi_3 \rmd \xi_1)\,.
\end{equation}
The $\U(1)\times \U(1)$ isometry of the solution is generated by translations along the angles $\xi_{1}$ and $\xi_{2}$. Later on we will determine which of the two $\U(1)$'s is mapped to the R-symmetry and which one to the flavor symmetry in the QFT. The ranges of the coordinates are
\begin{equation}		
	(\theta,\xi_3) \in [0,\pi/2]\,,\qquad 	(\phi,\xi_1,\xi_2) \in [0,2\pi]\,.
\end{equation}
We now introduce the following functions 
\begin{equation}\label{Eq: functions needed for the uplift}
	\begin{aligned}
		X = &\,\sqrt{2-Y^2}\,,\\ 
		a_1^\pm =&\, Y^2(2\pm X \cos 2\phi)\,,\\
		a_2^\pm =&\, X(\cos 2\xi_3 +(1 \pm X) (\cos 2\xi_3 - X \cos 2\phi) \sin^2\theta )\,,\\
		a_3^\pm =&\, (2 \pm X \cos 2\xi_3) \cos^2\theta \mp X^2\cos 2\phi \sin^2\theta - 4\,,\\
		b^\pm_X =&\, (2\pm X)(2+X\cos 2\xi_3)\,,\\
		v_0 =&\,-\sin 2\theta \cos\theta \sin \phi\,,\\
		v_1 =&\, X(X\cos 2\xi_3 \cos^2\theta + \cos 2\phi \sin^2\theta)\,,\\
		v_2 =&\, X(\cos 2\xi_3 \cos^2\theta + XY^2\cos 2\phi \sin^2\theta) \,.
	\end{aligned}
\end{equation}
The full 10d metric in Einstein frame is given by
\begin{equation}\label{Eq: Full 10d metric in Einstein}
	\dd s_{10}^2 =  \frac{\Delta^{1/4}}{2 Y^{2/3}}(\dd s_5^2 + Y^{2/3} \dd \tilde{s}^2_{S^5})
\end{equation}
where $\dd s_{5}^2$ is given in \eqref{Eq: five-dimensional metric}, the warp factor reads
\begin{equation}
\begin{aligned}
	\Delta =& 16\big(4(\cos^2\theta + Y^2 \sin^2\theta) - \cos^2\theta (v_1 \cos 2\xi_3 - v_2 \cos 2\phi ) -v_2 \cos 2\phi \big)\,,
\end{aligned}
\end{equation}
and finally the metric on the internal squashed five-sphere is given by
\begin{equation}
\begin{aligned}
	\dd \tilde{s}_5^2 =& \frac{64}{g^2 \Delta}\Big[ \tfrac12 Y^2 \cos^2\theta \Big((1+\sin^2\theta) \frac12\left[\dd\xi_1^2 +\dd\xi_2^2 +2 \cos 2\xi_3 \dd\xi_1 \dd\xi_2\right] \\
	&-X \cos 2\phi \sin^2\theta \frac12\left[\cos 2\xi_3 (\dd\xi_1^2 + \dd\xi_2^2) + 2\dd\xi_1 \dd\xi_2 \right] \\
	&+ \frac14 \sin^2 2\xi_3 \cos^2\theta \left[\rmd \xi_1^2 + \rmd \xi_2^2 (\frac{4}{Y^2}-1) \right]\Big)\\
	& + 2\cos^2\theta\Big( (X^2 + (1-X^2)\cos^2 \xi_3)\cos^2\theta + Y^2 (1+ \tfrac{1}{4X}\cos 2\xi_3 (a_3^--a_3^+)) \Big)\dd\xi_3^2\\
	& + \sin^2\theta \left( 4\cos^2 \phi (1+\sin^2\theta) + a_3^+ \cos 2\phi \right)\dd\phi^2 \\
	& + X \sin 2\xi_3 \sin2\phi \sin^2 2\theta \,\dd\xi_3 \dd\phi \\
	& + \Big( 2+\left(\frac12\cos 2\phi-\frac{2}{X}\cos 2\xi_3\right)\cos^2\theta(a_3^--a_3^+) + 2(1-X^2)\sin^2\theta \\
	& - \cos 2\xi_3 \left[\frac{1}{X} v_2 \sin^2\theta+\cos^2\theta \cos 2\xi_3 (4\cos^2\theta - (1+X^2) \sin^2\theta) \right] \Big)\dd\theta^2  \\
	& - \frac{X}{1-X^2}\sin 2\theta \Big[\sin 2\xi_3 (X v_1 - v_2 + 2(1-X^2) \cos 2\phi) \dd \xi_3 \\
	& - \sin2\phi (v_1 - \frac{1}{X} v_2 +2(1-X^2) \cos 2\xi_3) \dd\phi\Big] \dd\theta   \Big]\,.
\end{aligned}
\end{equation}
The complex axio-dilaton reads
\begin{equation}\label{eq:axiodildef}
\tau = C_0 +\rmi \rme^{-\Phi} = \frac{ \Lambda_{22} \hat \tau + \Lambda_{21}}{\Lambda_{12} \hat \tau + \Lambda_{11}}\,,
\end{equation}
where
\begin{equation}
\hat \tau = \frac{(1 - X^2)\sin2\phi \sin^2\theta + \rmi \sqrt{\Delta}/4}{a_3^+ + 2(X^2+1) \cos^2 \phi \sin^2\theta} \rme^{-2\varphi}\,,\quad \Lambda =
\frac{\rme^{-T/2}}{\sqrt{2\sinh T}}\left(
 \begin{array}{cc}
 	{\rme^{T}} & -1\\
 	-1 & {\rme^{T}}
 \end{array}\right)\,.
\end{equation}
Using the same $\SL(2,\mathbf R)$ element, the RR and NSNS two-forms are given by
\begin{equation}
	\left(\begin{array}{c}
		B_2 \\
		C_2
	\end{array}\right) = \Lambda \cdot \left(\begin{array}{c}
	\hat B_2 \\
	\hat C_2
\end{array}\right)\,,
\end{equation}
with
\begin{equation}
	\begin{aligned}
		\hat B_2 =& \frac{16 \rme^{\varphi}}{g^2\Delta}\partial_{\phi} (v_0)  \Big[  \frac14 \sin 2\xi_3 [(1+X)(b^+_{-X} \cos^2\theta + a_1^- \sin^2\theta)\dd(\xi_1+\xi_2) \\
		&\hspace{3.9cm} + (1-X)(b_{-X}^- \cos^2\theta + a_1^+ \sin^2\theta)\dd(\xi_1-\xi_2)]\wedge \dd \xi_3\\
		&\hspace{1.5cm} - \frac12\tan\phi [-(1+X)\cos^2 \xi_3 (a_3^+ + 4 X \cos^2 \phi \sin^2\theta) \dd (\xi_1+\xi_2) \\
		&\hspace{3.3cm}+ (1-X)\sin^2 \xi_3 (a_3^+ - 4 X \cos^2 \phi \sin^2\theta) \dd (\xi_1-\xi_2)]\wedge \dd \phi\\
		&\hspace{1.5cm} - \frac{1}{\sin 2\theta} [-(1+X)\cos^2 \xi_3 (a_1^- \sin^2\theta + (2-a_2^-) \cos^2\theta)\dd(\xi_1+\xi_2) \\
		&\hspace{3cm}+ (1-X)\sin^2\xi_3 (a_1^+ \sin^2\theta + (2-a_2^+)\cos^2\theta)\dd(\xi_1-\xi_2)]\wedge \dd\theta \Big]\,,
	\end{aligned}
\end{equation}
\begin{equation}
\begin{aligned}
	\hat C_2 =& \frac{16 \rme^{-\varphi}}{g^2\Delta} v_0  \Big[ - \frac14 \sin 2\xi_3 [(1-X)(b^+_{X} \cos^2\theta + a_1^- \sin^2\theta)\dd(\xi_1+\xi_2) \\
	&\hspace{3.3cm} + (1+X)(b_{X}^- \cos^2\theta + a_1^+ \sin^2\theta)\dd(\xi_1-\xi_2)]\wedge \dd \xi_3 \\
	&\hspace{1.5cm} + \frac12 \cot\phi [(1-X)\cos^2 \xi_3 (a_3^- - 4 X \cos^2 \phi \sin^2\theta) \dd (\xi_1+\xi_2) \\
	&\hspace{3cm}- (1+X)\sin^2 \theta_3 (a_3^- + 4 X \cos^2 \phi \sin^2\theta) \dd (\xi_1-\xi_2)]\wedge \dd \phi \\
	&\hspace{1.5cm}+\frac{1}{\sin 2\theta} [-(1-X)\cos^2 \xi_3 (a_1^- \sin^2\theta + (2+a_2^+) \cos^2\theta)\dd (\xi_1+\xi_2) \\
	&\hspace{3cm}+ (1+X)\sin^2\xi_3 (a_1^+ \sin^2\theta + (2+a_2^-)\cos^2\theta)\dd(\xi_1-\xi_2)]\wedge \dd\theta \Big]\,.
\end{aligned}
\end{equation}
Finally, the four-form with legs along the internal five-sphere is given by 
\begin{equation}
	\begin{aligned}
		C_4^\text{int} =& \frac{64}{g^4}\Big[\frac{\cos^2\theta\sin 2\xi_3\sin 2\theta}{ \Delta} \Big((-v_1 \sin 2\xi_3 \dd \phi + v_2 \sin2\phi \dd \xi_3 ) \wedge \dd \theta\\
		&\, + \frac14 \sin 2\theta \left(2(Y^2-1) - (v_1\cos 2\xi_3 -v_2 \cos 2\phi) \dd \xi_3 \wedge \dd\phi \right)\wedge \dd \xi_1 \wedge \dd\xi_2 \Big) +\omega_4\Big]\,,
	\end{aligned}
\end{equation}
where $\omega_4$ is the integrated volume form on $S^5$
\be
\omega_4=\frac{1}{16} \sin (2\xi_3)  \cos^4 \theta 	\rmd \xi_1 \wedge \rmd \xi_2 \wedge  \rmd \xi_3 \wedge \rmd \phi\,.
\ee

To relate the 5d gauge coupling $g$ to the number of D3-branes sourcing the 10d background we compute the integral of $F_5$ to find that
\begin{equation}\label{Eq: Number of D3 branes}
	N = \int_{S^5} \left. F_5 \right|_{S^5} = \int_{S^5} \rmd C_4^\text{int} =  \frac{4}{\pi g^4 \ell_s^4 }\,,
\end{equation}
where we projected the 5-form to its $S^5$ components. For convenience we also provide the four-form gauge potential with legs along the AdS$_4$ space. This can be found using the self-duality constraint in type IIB supergravity
%
\begin{equation}
	F_5 = \star F_5\,.
\end{equation}
Explicitly, working out the Hodge-dual of the five-form flux on the internal space and integrating over the $\varphi$ direction one finds that the four-form along the AdS space is given by
\begin{equation}
	C_4^{\text{AdS}_4} = \frac{4}{g^4} \left( 6\varphi + 2 X \cos 2\xi_3 \cos^2\theta +(1+X^2) \cos 2\phi \sin^2\theta \right) \vol_{\text{AdS}_4}\,.
\end{equation}
One can show that 10d background presented above solves the type IIB equations of motion, for which we have used the conventions of \cite{Bobev:2018hbq}.

Since this background is rather involved we present an additional consistency check by studying the flavor symmetry enhanced point $Y=\sqrt{2}$ which can be compared to the 10d solution presented in \cite{Bobev:2020fon}. Indeed, taking $Y = \sqrt{2}$ the background presented above greatly simplifies. The metric becomes\footnote{Here we correct a mistake in \cite{Bobev:2020fon}. To get the correct expression presented here one should perform the rescaling $g = 2^{\frac{1}{32}} g_\text{there} $.}
\begin{equation}\label{eq:metSU2Jfold}
	\rmd s_{10}^2 = \frac{2 (4 w)^{1/4}}{g^2} \left( \rmd \varphi^2 + \rmd s_{\text{AdS}_4}^2 + \rmd \theta^2 + \sin^2\theta \rmd \phi^2 + \cos^2\theta \left( \sigma_3^2 + \frac{\sigma_1^2 + \sigma_2^2}{w} \right) \right)\,,
\end{equation}
with the $\mathfrak{su}(2)$ one-forms defined in \eqref{Eq: su2 one forms}, and $w = 1 + \sin^2\theta$. The form potentials on the internal space equal 
\begin{equation}\label{eq:formsSU2Jfold}
	\begin{aligned}
		\hat C_2 =&\, \frac{2 \sqrt{2} \rme^{-\varphi}}{g^2} \frac{\cos \theta \sin \phi}{w} \left( \sin 2\theta \sigma_1 \wedge \sigma_2 - w (\rmd \theta - \frac12 \cot \phi \sin2\phi \rmd \phi) \wedge \sigma_3 \right)\,,\\
		\hat B_2 =&\, \frac{2 \sqrt{2} \rme^{-\varphi}}{g^2} \frac{\cos \phi \cos \theta}{w} \left( \sin 2\theta \sigma_1 \wedge \sigma_2  - w (\rmd \theta - \frac12 \tan \phi \sin 2\theta  \rmd \phi) \wedge \sigma_3 \right)\,,\\
		C_4 =&\, -\frac{2}{g^4 w} \cos^2\theta \sin^2 2\theta\, \rmd \phi \wedge \sigma_1 \wedge \sigma_2 \wedge \sigma_3\,. 
	\end{aligned}
\end{equation}
Finally, the axion and dilaton equal
\begin{equation}\label{eq:adSU2Jfold}
	\hat C_0  + \rmi \rme^{-\hat \Phi} = \rme^{-2\varphi}\frac{\rmi \sqrt{1+\sin^2\theta} + \frac12 \sin2\phi \sin^2\theta}{1+\sin^2\phi \sin^2\theta}\,.
\end{equation}
These expressions agree with the background presented in \cite{Bobev:2020fon}.%

\subsection{Three parameter family in type IIB}\label{Sec: Two parameter family of novel type IIB vacua}
So far we described the type IIB background dual to a one-dimensional slice of the two-dimensional supersymmetric conformal manifold.\footnote{In Appendix~\ref{App: non-susy sols} we show that the full parameter space of AdS$_4$ vacua is four-dimensional, of which only a two-dimensional subspace is supersymmetric. In this section we construct a three-dimensional subspace of vacua directly in type IIB supergravity.} In \cite{Guarino:2021hrc} it was shown how one can turn on a special class of exactly massless scalar fields additional in the four-dimensional $\SO(6) \times \SO(1,1)$ gauged supergravity truncation. In the five-dimensional gauged supergravity description these massless scalars arise from Wilson loop deformations turned on along $S^1_{\mathfrak J}$ \cite{Berman:2021ynm}, breaking the supergravity gauge symmetry to its Cartan subalgebra. Since our supergravity solutions preserve $\U(1)_R \times \U(1)_F$ symmetry there are two such Wilson loops we can turn on. In type IIB supergravity these deformations are simply described by the coordinate transformation
\begin{equation}\label{Eq: chi defos}
	\xi_1 \rightarrow \xi_1 + \chi_1 \varphi\,,\quad \xi_2 \rightarrow \xi_2 + \chi_2 \varphi\,,
\end{equation}
which is well-defined only locally due to the $\SL(2,\mathbf{Z})$ monodromy along $S^1_{\mathfrak J}$. One can show that such a deformations applied to the type IIB supergravity background described in the previous section indeed solves the equations of motion. For generic values of $\chi_i$ the solution breaks all supersymmetry, and thus we find a holographic description of a three-dimensional conformal manifold of non-supersymmetric CFTs. This raises the important question whether the corresponding AdS$_4$ vacua are stable. In \cite{Giambrone:2021wsm} a two-dimensional non-supersymmetric slice of this three-dimensional family of vacua was studied, and it was argued that indeed they are perturbatively stable. Furthermore, it was argued that neither higher derivative corrections in the gravitational theory, nor known non-perturbative effects can spoil the stability of the vacua. Much of the reasoning presented in \cite{Giambrone:2021wsm} relies on the fact that the non-supersymmetric vacua are continuously connected through diffeomorphisms to the $\mathcal N=4$ preserving vacuum, which in our coordinates is found at $(Y,\chi_1,\chi_2) = (1,0,0)$. Since all the type IIB vacua we presented above are connected to the $\mathcal N=4$ supersymmetric one, in a similar fashion, we find that the results in \cite{Giambrone:2021wsm} imply that our full three-dimensional family of type IIB vacua is stable as well, regardless of whether supersymmetry is broken. We leave a more detailed and rigorous analysis of the very interesting question of the stability of these vacua for future work.

The $\mathcal{N}=2$ supersymmetric AdS$_4$ vacua are of special interest. For these solutions we can compute physical  observables and hope to compare to known exact QFT results obtained via supersymmetric localization. This will be of particular interest in the following sections and thus we now study which two-dimensional subset of vacua preserve at least $\mathcal N=2$ supersymmetry. To preserve supersymmetry only a linear combination of the $\chi_i$ parameters that leaves the $\U(1)_R$ isometry unchanged can be turned on. To determine this linear combination one can simply take the limit $Y\rightarrow \sqrt{2}$, in which case the background corresponds to the one described in Section 3.3 in \cite{Bobev:2020fon}, which has an isometry enhancement to $\SU(2)_F\times\U(1)_R$. The full solution in this limit is given in \eqref{eq:metSU2Jfold}, \eqref{eq:formsSU2Jfold}, and \eqref{eq:adSU2Jfold} above. The $\SU(2)_F$ isometry is described by the left invariant one-forms, the $\U(1)_F \subset \SU(2)_F$ preserved away from the enhanced point is generated by $\rmd \xi_1$. The $\U(1)_R$ on the other hand is described by a rotation between $\sigma_1 $ and $\sigma_2$, which is in turn described by a shift in $\xi_2  $. We thus see that the $\chi_i$ deformation preserving supersymmetry, leaving the $\U(1)_R$ untouched, is along $\rmd \xi_1$ in equation \eqref{Eq: chi defos}. We therefore conclude that the deformation parametrized by $\chi_1 $ in \eqref{Eq: chi defos} preserves $\mathcal{N}=2$ supersymmetry. In summary, the 10d uplift of the two-parameter family of $\mathcal{N}=2$ supersymmetric AdS$_4$ vacua found in \cite{Bobev:2021yya} is given by the 10d background above parametrized by $Y$ and $\chi_1$ with vanishing $\chi_2$.

\subsection{Infinite distances on the conformal manifold}\label{Sec: Infinite distances on the conformal manifold}
As mentioned in the introduction, previous holographic studies of the $\mathcal N=2$ conformal manifold of S-folds showed the manifold to be non-compact. The Zamolodchikov metric of the manifold was computed to equal \cite{Bobev:2021yya}
\begin{equation}
	\dd s_Z^2 = \frac{4-Y^2}{2}\left(\frac{\dd Y^2}{Y^2(2-Y^2)} + d\chi^2\right)\,,
\end{equation}
which has unbounded geodesic curves. In this parametrization the infinite distance is reached when $Y\rightarrow 0$, and in previous studies it was shown that the five-dimensional solution is badly singular in this limit since the full AdS$_4$ space vanishes \cite{Arav:2021gra}, which can be directly seen from equation \eqref{Eq: five-dimensional metric}.

In this section we study the behavior of the full ten-dimensional solution in the limit where $Y\rightarrow 0$. As can be directly seen from \eqref{Eq: Full 10d metric in Einstein} we find that the singular behavior of the metric is improved in ten dimensions. To leading order in a small $Y$ expansion the metric becomes 
\begin{equation}
	\begin{aligned}
		\dd s_{10}^2 = \frac{(\Delta^{(0)})^{1/4}}{g^2}\left[\left(2^{4/3} \dd \varphi^2\right)\frac{1}{Y^2} +\left( 2^{7/3} \dd s_{\text{AdS}_4}^2 +\frac{1}{2} (\dd s_5^2)^{(0)}\right) + \mathcal O(Y)\right]\,,
	\end{aligned}
\end{equation}
where $(\cdot)^{(p)}$ denotes the leading order non-trivial term in the expansion of $(\cdot)$, which is of order $Y^p$. It turns out that to zeroth order the remaining type IIB field content becomes independent of the parameter $Y$
\begin{equation} 
	\begin{aligned}
		&\rme^{\hat \Phi} = (\rme^{\hat \Phi})^{(0)} + \mathcal O(Y)\,,\quad \hat C_0 =\,(\hat C_0)^{(0)} + \mathcal O(Y)\,,\\
		&B_2 = (\hat B_2)^{(0)} + \mathcal O(Y)\,,\quad \hat C_2 = (\hat C_2)^{(0)} + \mathcal O(Y)\,,\quad C_4 = (C_4)^{(0)} + \mathcal O(Y)\,.
	\end{aligned}
\end{equation}
The full expressions for the leading order terms are rather involved functions depending on the internal coordinates, and we have summarized the details of their explicit form in Appendix~\ref{App: Y to zero background}. We thus find that to leading order in $Y$ the only dependence on $Y$ in the IIB background is given by the metric component that determines the radius of the S-fold circle $\rmd \varphi$. Importantly, as we take $Y \rightarrow 0$ the size of the circle blows up. This property of the 10d type IIB supergravity background may explain some of the features of the KK spectrum of this family of AdS$_4$ solutions that was computed in \cite{Cesaro:2021tna} using tools from exceptional field theory, see \cite{Malek:2019eaz} and references thereof. Indeed, it was found in \cite{Cesaro:2021tna} that the masses of certain unprotected KK modes do not increase as one takes $Y \rightarrow 0$, i.e. they stay ``light'' in this limit.

In view of the CFT distance conjecture discussed in \cite{Perlmutter:2020buo}, see also \cite{Baume:2023msm}, one may expect to find  a tower of massless higher-spin fields, corresponding to conserved higher-spin operators in the dual CFT, emerging in the spectrum as we take $Y \rightarrow 0$. We could not find any evidence for the existence of such modes and given the results above and the ones in \cite{Cesaro:2021tna}  it is quite clear that if these modes exist they should not arise from the supergravity fields but rather from excitations associated with fundamental strings or D-branes. It will be most interesting to understand whether the locus $Y \rightarrow 0$ provides a counterexample to the CFT distance conjecture and it does not to identify the origin of the infinite tower of light higher-spin modes.

Before closing this section we would like to note that the increase in the radius of the $\varphi$ circle in the metric as $Y \rightarrow 0$ may have an interesting interpretation in the dual SCFT. From \eqref{Eq: five-dimensional metric} and \eqref{Eq: Full 10d metric in Einstein} it is clear that changing $Y$ simply changes the size of the circle parametrized by $\varphi$. On the other hand, as discussed below \eqref{Eq: five-dimensional metric}, the periodicity of that same S-fold circle is given by $T$. In the dual 3d S-fold SCFT $Y$ and $T$ are mapped to the exactly marginal coupling $\lambda$ and CS level $k$, respectively. In the formal superpotential that specifies the gauging of the $T[U(N)]$ theory the dependence on these parameters is given by \cite{Bobev:2021yya}
\begin{equation}
	W_{\text{IR}} = -\frac{2\pi}{k} \Tr (\mu_H \mu_C) + \lambda \,\Tr (\mu_H\mu_C)\,.
\end{equation}
We therefore observe that a change in $\lambda$ can be reabsorbed by a change in the CS level $k$ in the superpotential. This analysis seems to suggest that at infinite distance on the conformal manifold, where $Y\rightarrow 0$, the SCFT may have an equivalent description as an S-fold SCFT with the superpotential above in the limit of infinite CS level $k$. It will be very interesting to study this analogy further and understand whether it can be made more precise.

\section{Non-perturbative corrections to the $S^3$ partition function}\label{Sec: non perturbative contributions}
As reviewed in the Introduction, the $S^3$ partition function of the $\mathcal{N}=4$ S-fold SCFT was computed in \cite{Assel:2018vtq,Ganor:2014pha,Terashima:2011qi} using supersymmetric localization. The result can be written as a matrix model of the following form\footnote{The $\sinh$ term one would expect from the vector multiplet contribution cancels agains a term in the $T[\U(N)]$ contribution, greatly simplifying the matrix model. Note also that the same matrix model has appeared in \cite{Matsuura:2022dzl} in a different context.}
\begin{equation}\label{Eq: Partition function of Sfold}
\begin{aligned}
	\mathcal Z_N =& \,\frac{1}{N!} \int \rmd^N \sigma \, Z_{\text CS}(\sigma) Z_{\text vec}(\sigma) Z_{T[\U(N)]}(\sigma,-\sigma)\\
	=& \frac{1}{N!} \sum\limits_{\tau \in S_N} (-1)^\tau \int \rmd^N \sigma\, \rme^{i \pi k \sum_{j=1}^N \sigma_j^2 \,- \, 2\pi i \sum_{j=1}^N \sigma_j \sigma_{\tau(j)}}\,,
\end{aligned}
\end{equation}
where $k$ is the Chern-Simons level. Defining the quantities
\begin{equation}\label{eq:qSfolddef}
	q = \rme^{-T}\, \qquad \text{and} \quad k = 2 \cosh T\,,
\end{equation}
this matrix model can be solved analytically for any $N$ and $k$ to find
\begin{equation}\label{Eq: susy partition function S-fold}
	\mathcal Z_N = \frac{q^{N^2/2}}{(q)_N}\,,
\end{equation}
where $(q)_N$ is the $q$-Pochhammer symbol 
\begin{equation}
	(q)_N = \prod\limits_{m=1}^N (1-q^m)\,.
\end{equation}
Taking the large $N$ limit we can find the leading order term in the $S^3$ free energy 
\begin{equation}\label{Eq: leading supergravity action}
	F_{S^3} = - \log \mathcal Z_N  = \frac{N^2 T}{2}+\ldots\,.
\end{equation}
This result agrees with the on-shell action of the dual AdS$_4$ solution, see \eqref{eq:SonshellAdS4}. Moreover, in 3d SCFTs exactly marginal deformations are exact with respect to the localization supercharged used for the calculation of the $S^3$ partition function and therefore, as discussed in \cite{Bobev:2020fon,Bobev:2021yya}, the result in \eqref{Eq: leading supergravity action} is valid along the full two-dimensional conformal manifold. 

Inspired by some recent observations in \cite{Imamura:2021ytr,Gaiotto:2021xce,Murthy:2022ien} we can proceed to rewrite the $S^3$ partition function \eqref{Eq: susy partition function S-fold} in a suggestive way. First we note that the infinite extension of the Pochhammer symbol $(x;q)_\infty = \prod\limits_{n=0}^\infty (1- x q^n)$ can be used to derive the following identity
\begin{equation}
 \frac{(q)_\infty}{(q)_N} = (q^{N+1};q)_\infty\,.
\end{equation}
Importantly, the right hand side above can be rewritten as a sum using the relation \cite{Zagier2007}
\begin{equation}
	(x;q)_\infty = \sum\limits_{n=0}^\infty  \frac{(-1)^n q^{n(n-1)/2}}{(q)_n} x^n\,.
\end{equation}
Using the identities above it is easy to show that the $S^3$ path integral of the S-fold SCFT of interest \eqref{Eq: susy partition function S-fold} can be written as
\begin{equation}\label{Eq: index interpretation}
	\mathcal{Z}_N = \frac{q^{N^2/2}}{(q)_{\infty}}\left( 1 + \sum\limits_{n=1}^N (-1)^n\frac{q^{n(n+1)/2}}{(q)_n}q^{n N}\right)\,, 
\end{equation}
or equivalently 
\begin{equation}\label{Eq: finite corrections to partition function}
	\mathcal{Z}_N = \frac{q^{N^2/2}}{(q)_{\infty}}\left(1 + \sum\limits_{n=1}^\infty (-1)^n \mathcal Z_n q^{n(N+1/2)}\right)\,.
\end{equation}
This is a rather intriguing result since we find that the finite $N$ partition $S^3$ partition function can be written as an infinite series of terms each of which is determined again by the same supersymmetric partition function $\mathcal Z_n$. Such a structure is reminiscent of the giant graviton expansion \cite{Biswas:2006tj,Arai:2019xmp,Imamura:2021ytr,Gaiotto:2021xce} discussed in the context of the superconformal index, or $S^3 \times S^1$ partition function, of 4d $\mathcal{N}=1$ SCFTs. Remarkably, the expansion in \eqref{Eq: index interpretation} is identical to the non-perturbative series of giant gravitons in the $1/2$-BPS superconformal index of ${\rm U}(N)$ $\mathcal N=4$ SYM, see equation (1.9) in \cite{Gaiotto:2021xce}). The difference between the two expressions comes from the fact that the fugacity in the $\mathcal N=4$ index is replaced by the parameter $q$ in \eqref{eq:qSfolddef} determined by the parameter $T$ that controls the periodicity of the S-fold circle, or equivalently the CS level of the 3d SCFT. It is tempting to ponder the reason for this close similarity between two seemingly different supersymmetric observables. A possible explanation may arise through a proper definition of a new type of 4d supersymmetric index computed by the path integral of $\mathcal N=4$ SYM on $S^3 \times S^1_\mathfrak{J}$, where the notation $S^1_\mathfrak{J}$ indicates that one should perform an S-duality twist determined by $\mathfrak{J}$ in \eqref{eq:Jkmatrixdef} as one goes around the circle. One could then try to find a relation between the ordinary superconformal index defined as a path integral on $S^3\times S^1$ and the S-duality twisted index that will explain the relation above. Indeed, the fact that in the sum \eqref{Eq: index interpretation} the numerator contains only integer powers of $q$ resembles the structure of an index. Moreover, note that the $q^{N^2/2}$ prefactor on the right hand side in \eqref{Eq: finite corrections to partition function} resembles the supersymmetric Casimir energy on $S^3\times S^1$ which is determined by the conformal anomalies of the 4d SCFT, see \cite{Assel:2015nca,Bobev:2015kza}. Stripping off this prefactor then naturally leads to the conjecture that the factor of $1/(q)_{\infty}$ should be interpreted as the supergravity contribution to the putative S-duality twisted index. It will be most interesting to explore these questions further and study the existence and properties of such S-duality twisted indices of 4d SCFTs.

Regardless of the existence of a properly defined S-duality twisted index we can adopt a string theory perspective to shed more light on the structure of the partition function in \eqref{Eq: finite corrections to partition function}. The form of this expression suggests that in string theory this partition function can be thought of as  a leading order supergravity piece coming from the large $N$ stack of D3-branes wrapping $S^3\times S^1_{\mathfrak J}$, accompanied by an infinite sum of non-perturbative corrections coming from D3-branes wrapping $\tilde S^3 \times S^1_\mathfrak{J}$. Notably, the two three spheres are embedded differently in the type IIB supergravity background. The $S^3$ is simply given by the conformal boundary of the Euclidean AdS$_4$ while the $\tilde S^3$ is embedded in the squashed internal $S^5$ spanned by the coordinates $(\theta,\phi,\xi_{1,2,3})$ used in Section~\ref{Sec: The novel type IIB S-fold backgrounds}. The explicit 10d supergravity solution constructed in Section~\ref{Sec: The novel type IIB S-fold backgrounds} indeed allows us to check this intuition in detail.

To this end let us study a probe Euclidean D3-brane that  wraps the three angular directions parametrized by $\xi_{1,2,3}$, and the S-fold circle parameterized by $\varphi$. 
To compute the on-shell action of this D3-brane we first have to minimize the brane action that consists of the DBI action, coupling the brane to the NS-NS sector, and the WZ terms that determine the coupling to the RR forms. It turns out that for the D3-brane configurations described above it is sufficient to minimize the DBI action, and furthermore that the RR forms and the $B_2$-field do not contribute either to the extremization problem or to the on-shell action of the brane. With this in mind the D3-brane action is simply controlled only by the Einstein frame metric and reads
\begin{equation}
S_{\text{D3}} = \frac{2\pi}{(2\pi \ell_s)^4} \int \rmd^4 \sigma\,  \sqrt{\det G_{ab} }\,.
\end{equation}
Using static gauge we simply have to extremize this action with respect to the transverse coordinates, which shows that in the internal space the probe D3-brane sits at the points
\begin{equation}
	\sin2 \theta = 0\,,\quad \text{and}\quad  \sin 2\phi = 0\,.
\end{equation}
The pulled back metric on the D3-brane in this configuration becomes\footnote{The $\text{SU}(2)$ one-forms are defined in \eqref{Eq: su2 one forms}.}
\begin{equation}\label{Eq: D3 brane metric}
\begin{aligned}
	\rmd s_{\text{D3}}^2 =& \frac{4\Delta_{\text{D3}}^{1/4}}{g^2}\Big( \frac{\rmd \varphi^2}{Y^2} + \sigma_1^2 +\sigma_2^2 + \frac{1}{\Delta_{\text{D3}}} \left(2 + (2-Y^2) (2-\cos^2\xi_3)  \right)\sigma_3^2\\
	 &\qquad - \frac{2-Y^2}{2\Delta_{\text{D3}}} \left(   \rmd\xi_1^2 +\rmd\xi_2^2 +(2+\sin^2 2\xi_3) \cos2\xi_3 \,\rmd \xi_1 \rmd \xi_2\right) \Big)\,,
\end{aligned}
\end{equation}
where
\begin{equation}
	\Delta_{\text{D3}} = 4 - (2 - Y^2) \cos^2 2\xi_3\,.
\end{equation}
The on-shell action can be easily calculated to find the simple result
\begin{equation}\label{eq:SD3onshell}
	S_{\text{D3}}  = \frac{4T}{\pi g^4 \ell_s^4} = T N\,,
\end{equation}
where we have used \eqref{Eq: Number of D3 branes} to relate $g$ and $\ell_s$ to the number $N$ of D3-branes that produce the AdS$_4$ supergravity solution. We would like to stress that this probe brane analysis is valid for any value of $Y$, i.e. for the full family of AdS$_4$ supergravity solutions discussed in Section~\ref{Sec: The novel type IIB S-fold backgrounds}. This is compatible with the expectation that the probe D3-brane with the embedding described above preserves supersymmetry. 

Given the probe brane analysis above, one should expect that the non-perturbative contributions of such probe D3-branes to the full string theory partition function on the AdS$_4$ S-fold background takes the form
\begin{equation}\label{Eq: non-perturbative corrections string partition function}
	\mathcal Z_{\text{np}} = \sum\limits_{n=0}^\infty Z^{\text{D3}}_n \rme^{-n S_{\text{D3}}} = \sum\limits_{n=0}^\infty Z^{\text{D3}}_n q^{n N}\,,
\end{equation}
where in the second identity we have used \eqref{eq:qSfolddef} and \eqref{eq:SD3onshell}. This result is in remarkable agreement with the $S^3$ partition function of the dual SCFT as presented in \eqref{Eq: index interpretation} and \eqref{Eq: finite corrections to partition function}. The off-set by $1/2$ in the $q^{n(N+1/2)}$ term of \eqref{Eq: finite corrections to partition function} could perhaps be explained by a finite $N$ correction to the Newton's constant, similar to what was found in \cite{Bergman:2009zh}. Nevertheless, as noted above, this shift in $N$ is crucial when interpreting the results in \eqref{Eq: index interpretation} as a putative 4d SCFT index.

Note that the D3 brane metric in \eqref{Eq: D3 brane metric} is topologically that of $S^1 \times S^3$, however the three sphere is equipped with a squashed metric. It is thus rather remarkable that the non-perturbative corrections the probe D3-branes contribute to the full string partition function takes the same form as the one of the $N$ backreacted D3-branes that produce the supergravity background since these $N$ branes wrap a round sphere. A potential explanation for this fact could be that we are computing a supersymmetric quantity and that the squashing deformation of the $S^3$ are $Q$-exact. A hint in support of this interpretation is provided by the observation that the $S^3$ squashing is controlled by the parameter $Y$, which is dual to an exactly marginal $Q$-exact deformation of the S-fold partition function. 

It will be interesting to use string theory to reproduce the $ Z_n^{\text{D3}}$ contribution in \eqref{Eq: non-perturbative corrections string partition function}. One way to do this is to study small fluctuations of the fields on the worldvolume of the D3-brane and compute their superdeterminants. This calculation may proceed similarly to what was done recently for the fundamental string wrapping Euclidean two-cycles in various holographic backgrounds in \cite{Gautason:2023igo}, which was subsequently generalized to M-theory and fluctuating M2-branes in \cite{Beccaria:2023ujc,Beccaria:2023sph}. The D3-brane one-loop partition function has been computed with such methods in a different setup in \cite{Buchbinder:2014nia}. Another way to go about this may be to compute the $\tilde S^3$ partition function of the S-fold theory which lives on the Euclidean probe D3-brane wrapped in the internal $\tilde{S}^3$, similarly to what was done in \cite{Arai:2019xmp} for Euclidean D3-branes wrapping an $S^3$ in the internal space of AdS$_5 \times S^5$ exploiting the giant graviton expansion in the bulk. It will be very interesting to explore this question further.

\section{Holographic line operators}\label{sec:hololines}
In this section we will use the new ten-dimensional background presented in Section~\ref{Sec: The novel type IIB S-fold backgrounds} to compute the expectation values of line operators in the dual CFT. The CFT line operators we consider are dual to fundamental strings, D1-branes, or more general $(p,q)$-strings wrapping the equator of $S^3$ and extending into the bulk geometry. We also study the vacuum expectation value of the Wilson loop using the results from supersymmetric localization in the dual 3d CFT and find good schematic agreement with the supergravity analysis.

\subsection{Gravity}
We consider the on-shell action of a $(p,q)$-string that wraps the equator of $S^3$ on the boundary of Euclidean AdS$_4$. The string extends into the bulk geometry in such a way that its worldsheet spans an $\text{AdS}_2$ subspace of  $\text{AdS}_4$. We assume that the string sits at a fixed location along the remaining directions of the 10d background. This is done in order to retain the isometries of AdS$_2$ as symmetries of the string. The classical action of the $(p,q)$-string takes the form 
\begin{equation}\label{Eq: 1-brane action}
	S_{(p,q)} = \frac{1}{2\pi \ell_s^2} \int \left[ \rmd^2 \sigma \sqrt{ \mathbf{p} \cdot \mathcal M \cdot \mathbf{p}^T\, \det[G_{ab}] } +i \mathbf{p} \cdot A_2 \right] \,,
\end{equation}
where $\mathbf{p} = (p,q)$ determines the charge of the string, $A_2 = (C_2,B_2)$ gives the coupling to the background type IIB two-forms, and the matrix $\mathcal M$ encodes the axio-dilaton, see \eqref{eq:axiodildef}
\begin{equation}
	\mathcal M = \rme^{\Phi} \left( \begin{array}{cc}
		\tau \bar{\tau} & C_0 \\
		C_0 & 1
	\end{array} \right)\,.
\end{equation}
The fundamental string has charge $\mathbf p = (0,1)$, which corresponds to a Wilson line in the fundamental representation in the dual CFT, while any other charge corresponds to 't~Hooft or dyonic line operators in the QFT. Extremizing the on-shell action of the $(p,q)$-string leads to a number of different solutions. Notably, only one of them is independent of the supergravity parameter $Y$ that is dual to the marginal coupling in the dual SCFT. Since the expectation value of a supersymmetric Wilson loop operators in 3d $\mathcal{N}=2$ theories are independent of exactly marginal couplings we expect that the on-shell action of the supersymmetric string should be independent of $Y$. We therefore assume that the other extrema for the on-shell action of the probe string break supersymmetry and we do not consider them further. The extremum we are interested in satisfies
\begin{equation}
	\cos \theta = 1 \,,\quad \cos 2\xi_3 = 0\,, \quad \rme^{2\varphi-T} = -\frac{p + \rme^{-T}q}{p + \rme^{T}q}\,.
\end{equation}
The on-shell action of such a $(p,q)$-string is easily computed after using the fact that the regularized volume of the unit-radius Euclidean AdS$_2$ is $-2\pi$. Using \eqref{Eq: Number of D3 branes} and the holographic dictionary we find that the expectation value of the $(p,q)$ line operator scales reads
\begin{equation}\label{pqvev}
\log{\cal W}_{(p,q)} = - S_{(p,q), {\rm on-shell}}^1 = \sqrt{\frac{2\pi N}{\sinh T} (p^2 + q^2 - p q \cosh T)}\,.
\end{equation}
In particular for the fundamental Wilson loop we find
\be\label{WLvev}
\log{\cal W} = \sqrt{\frac{2\pi N}{\sinh T}}\,.
\ee
The results \eqref{pqvev} and \eqref{WLvev} should be viewed as predictions for the holographic 3d $\mathcal{N}=2$ SCFT. Although we have not been able to independently and rigorously compute these line operator expectation values from the dual SCFT, we nevertheless were able to estimate the WL vev using the matrix model obtained by supersymmetric localization and the Wigner semi-circle law. We now proceed to briefly describe this calculation.

\subsection{Field theory}

To properly calculate the vev of a supersymmetric Wilson loop for the S-fold SCFT on $S^3$ one needs to apply supersymmetric localization and derive the analog of the matrix model result in \eqref{Eq: Partition function of Sfold} for the $S^3$ partition function of the SCFT. We will not pursue this calculation here but will rather draw on the results of \cite{Assel:2018vtq} and the calculation of the WL vev in the ABJM theory studied in \cite{Kapustin:2009kz} to make an educated guess about the result.

We will assume that the Wilson loop that preserves the large amount of supersymmetry wraps the great circle on $S^3$. Following \cite{Kapustin:2009kz} we can write this WL vev as\footnote{In this section we temporarily absorb a factor of $i$ in the definition of $\sigma^2$ as compared to the matrix model in \eqref{Eq: Partition function of Sfold} in order to follow standard Gaussian matrix model conventions.}
\begin{equation}\label{Eq: Wilson line vev}
	\langle W_R \rangle = \langle \text{Tr}_R \, \rme^{2\pi \sigma} \rangle\,,
\end{equation}
where $R$ is the representation of the Wilson loop with respect to the gauge group and $\sigma$ is same variable as in the matrix model in \eqref{Eq: Partition function of Sfold}. We hasten to add that this expression is a guess and should be treated with care especially since the S-fold SCFT of interest is non-Lagrangian. We proceed further with a crude approximation, namely we assume that in the large $N$ limit the eigenvalues $\sigma$  lay on a Wigner semi-circle distribution specified by the eigenvalue density
\begin{equation}\label{eq:rhosigmadef}
	\rho(\sigma) = \frac{2}{\pi \mu^2} \sqrt{\mu^2 - \sigma^2}\,, \quad \text{where} \quad -\mu <\sigma < \mu\,,
\end{equation}
where $\mu$ is a positive real number and we have normalized the eigenvalue density such that 
\begin{equation}
\int_{-\mu}^{\mu} \rho(\sigma) d\sigma =1\,.
\end{equation}
We stress that this assumption for the eigenvalue density is non-trivial since the matrix model at hand is not exactly Gaussian. With these assumptions at hand we can determine the value $\mu$ by computing the expectation value of $\sigma^2$ at large $N$ in two alternative ways. First, we can take two derivatives of the matrix model partition function in \eqref{Eq: Partition function of Sfold} with respect to the coupling. The role of coupling is played by the CS level $k$ which we promote temporarily to a continuous variable to find 
\begin{equation}
	\langle \sum\limits_j \sigma_j^2 \rangle = -\frac{1}{\pi} \frac{\partial}{\partial k} \log \mathcal Z_N[T] \approx \frac{N^2}{4\pi \sinh T}\,,
\end{equation}
where in the second relation on the right hand side we have taken the large $N$ limit. Alternatively we can first take the large $N$ limit and then use the eigenvalue density in \eqref{eq:rhosigmadef} to find
\begin{equation}
	\langle \sigma^2 \rangle = N \int\limits_{-\mu}^{\mu} \rmd \sigma \rho(\sigma) \sigma^2 = \frac{\mu^2 N}{4} \,,
\end{equation}
where the additional factor of $N$ arises from changing from a sum over eigenvalues to an integral. Comparing the two alternative approaches we conclude that
\begin{equation}
	\mu = \sqrt{\frac{N}{\pi \sinh T}}\,.
\end{equation}
To calculate the Wilson line vev in the fundamental representation we now simply have to evaluate the expectation value of $\rme^{2\pi \sigma}$ using the eigenvalue density $\rho(\sigma)$ and the result for $\mu$ above. The result is simple and reads
\begin{equation}
	\langle W_F \rangle = \int\limits_{-\mu}^{\mu} \rmd \sigma \rho(\sigma) \rme^{2\pi \sigma} = \frac{1}{2\mu} I_1(2 \pi \mu)\,,
\end{equation}
where $I_1$ is the modified Bessel function of the first kind. Taking a logarithm and using the large $N$ limit we find that
\begin{equation}\label{Eq: FT WL vev}
	\lim\limits_{N\rightarrow \infty}\log \langle W_F \rangle \approx 2 \pi \mu = \sqrt{\frac{4\pi N}{\sinh T}}\,.
\end{equation}
Despite the guesses we made this schematic calculation leads to a result that is of the same form as the one obtained by the probe string calculation on the gravity side and differs only by a simple numerical factor, see \eqref{WLvev}. We consider this to be a good qualitative check of holography in this setup. It will certainly be very nice to put this calculation of the WL vev on a more solid footing by a careful application of supersymmetric localization in the S-fold SCFT.

\section{4d $\mathcal{N}=2$ supergravity and S-fold black holes}
\label{sec:S-holes}

The goal of this section is to show that for each member of the family of supersymmetric AdS$_4$ vacua discussed in Section~\ref{Sec: The novel type IIB S-fold backgrounds} and Appendix~\ref{App: non-susy sols} there is a consistent truncation of the maximal 4d $\mathcal{N}=8$ $[\SO(6)\times \SO(1,1)] \ltimes \mathbf R^{12}$ gauged supergravity theory to the minimal 4d $\mathcal{N}=2$ gauged supergravity.

To restrict ourselves to the supersymmetric family of AdS$_4$ vacua we set
\begin{equation}
	z = -1 \,,\quad \chi_2 = 0\,.
\end{equation}
The scalar manifold is then spanned by only two real scalar fields, $x$ and $\chi_1$, that parametrize the coset element
\begin{equation}
	\mathcal V =\, \rme^{ \chi_1 \mathfrak h_1} \cdot \rme^{-\log \sqrt{2} \mathfrak t_1} \cdot \rme^{\f12\log\frac{3+x}{4} \mathfrak t_2} \cdot \rme^{\f{2}{x-3}\f{\sqrt{1-x}}{\sqrt{3+x}}[\mathfrak r_2,\mathfrak r_3]}\cdot \rme^{-\mathfrak r_2} \cdot \rme^{-\f{\sqrt{1-x}}{\sqrt{3+x}}\mathfrak r_3}\,.
\end{equation}
In addition to these two scalar fields the other bosonic fields in the truncation are the metric and two Abelian vector fields associated to the $\rm SO(6)$ generators $g_1$ and $g_2$ in \eqref{eq:defg1g2}. Using this information and the embedding tensor in \eqref{Eq: 4d embedding tensor} one can utilize the framework described in \cite{deWit:2007kvg} to write down the full Lagrangian of the truncated theory. To simplify the end result we define a complex scalar $\zeta$  such that
\begin{equation}
	x = \frac{8}{1+\cosh(2 \Re\,\zeta)} - 3\,,\quad \text{and} \quad  \chi_1 =\Im \, \zeta\,.
\end{equation}
The truncated bosonic Lagrangian then reads
\begin{equation}\label{Eq: Lagrangian 4d BH}
	\mathcal L =  R -  h_{IJ} F^I_{\mu\nu} F^{\mu\nu J} - (\partial_\zeta \partial_{\bar \zeta} \mathcal K) \zeta' \bar \zeta' + 6 \,,
\end{equation}
where the $F^I$ are two Maxwell fields and we have rescaled four-dimensional coordinates to absorb the length scale of the four-dimensional AdS vacuum $y\mapsto y L_\text{AdS} $ where $L_\text{AdS} = \sqrt{2}/g$. We have also defined
\begin{equation}
\begin{split}
	\mathcal K &= |\zeta|^2 - 2 \log \cosh \Re \,\zeta\,,\quad h = \left( \begin{matrix}{}
  1/4 & \\
  & \rho(\zeta,\bar \zeta)/2
\end{matrix}
 \right)\,, \\
	\rho(\zeta,\bar \zeta) &= \frac{1 + \cosh 2 \Re\, \zeta}{(1 + \cosh 2 \Re\, \zeta)\cosh 2 \Re\, \zeta + 8 (\Im\, \zeta)^2}\,.
\end{split}
\end{equation}
There are two important features of the Lagrangian in \eqref{Eq: Lagrangian 4d BH}. First, there is no potential for the scalar fields. This is compatible with the fact that they are exactly massless and reflects their role as gravitational duals to the exactly marginal couplings in the dual S-fold SCFT. Second, the kinetic term of the $F^{1}_{\mu\nu}$ Maxwell field is independent of the scalar fields. These properties of the truncated Lagrangian ensure that it is consistent with the equations of motion to set $F^{2}_{\mu\nu}=0$ and the scalars $x$ and $\chi_1$ to an allowed constant value that corresponds to any of the two-parameter family of supersymmetric AdS$_4$ vacua. Therefore, we conclude that for any value of $\zeta$ we have a consistent truncation to the following gravitational action
\begin{equation}\label{Eq: Lagrangian 4d BH}
S=-\frac{L^2_\text{AdS}}{16 \pi G_N}\int\star_4\Big(R - \frac{1}{4}  F_{\mu\nu}^1 F^{1\mu\nu} + 6\Big)\,.
\end{equation}
This is simply the action of the 4d Einstein-Maxwell theory with a negative cosmological constant which in turn is the bosonic sector of 4d $\mathcal{N}=2$ minimal gauged supergravity. The prefactor $L^2_\text{AdS}$ appears because of the coordinate rescaling performed above. The existence of this consistent supergravity truncation is compatible with the general expectation, see \cite{Gauntlett:2007ma,Bobev:2017uzs} and references thereof, that every supersymmetric $\mathcal{N}=2$ AdS$_4$ vacuum of string or M-theory should give rise to a consistent truncation to 4d $\mathcal{N}=2$ minimal gauged supergravity which describes holographically the universal dynamics of the energy-momentum tensor in the dual 3d $\mathcal{N}=2$ SCFT. The novel feature of the construction above is that since we have a two-parameter family of AdS$_4$ supersymmetric vacua we are led to a two-parameter space of consistent truncations to 4d $\mathcal{N}=2$ minimal gauged supergravity.

We can exploit the existence of this 4d $\mathcal{N}=2$ minimal gauged supergravity truncation in the context of holography and learn about some of the properties of the dual 3d $\mathcal{N}=2$ family of S-fold SCFTs. In particular, we find that the free energy of the SCFT placed on a compact smooth Euclidean 3-manifold $\mathcal{M}_3$,  $F_{\mathcal{M}_3}=-\log Z_{\mathcal{M}_3}$, can be written as
\begin{equation}
F_{\mathcal{M}_3} = S_\text{on-shell} \equiv \frac{\pi L^2_\text{AdS}}{2 G_N} \mathcal{F}\,.
\end{equation}
Here $\mathcal{F}$ is determined by the regularized on-shell action of an appropriate Euclidean gravitational solution of the equations of motion obtained from \eqref{Eq: Lagrangian 4d BH} which is given by a smooth filling of $\mathcal{M}_3$ with an accompanying Maxwell field. 
Using the same arguments as in \cite{Bobev:2021rtg} we can evaluate the prefactor to be $\pi L^2_\text{AdS}/2 G_N=N^2 T/2$.
Using this, we find the following prediction for the dual SCFT free energy on Euclidean 3-manifolds 
\begin{equation}
F_{\mathcal{M}_3} = \frac{N^2 T}{2} \mathcal{F}\,.
\end{equation}
There are a number of important Euclidean solutions of the 4d $\mathcal{N}=2$ minimal gauged supergravity theory that can be utilized in this context, see \cite{Bobev:2020egg,Bobev:2021oku} for a recent summary and a more extensive bibliography, together with the explicit form of the metric and Maxwell field as well as the regularized on-shell action $\mathcal{F}$. 

Below we briefly summarize the results for three examples of particular interest associated with well-known supersymmetric partition functions. Notably, the results below are independent of the exactly marginal couplings of the SCFT which suggests that these deformations should be $Q$-exact with respect to the localization supercharge.

\begin{itemize}

\item One can take $\mathcal{M}_3$ to be a particular squashed metric on $S^3$ that preserves supersymmetry. The corresponding supersymmetric AdS-Taub-NUT solution can be found in \cite{Bobev:2021oku}. Using the results above we then find that 3d S-fold SCFT free energy on this supersymmetric background  reads
\begin{equation}
F_{S^3_b} = \frac{(b+b^{-1})^2}{4}\frac{N^2 T}{2} \,,
\end{equation}
where the right hand side should be understood as the leading term in the large $N$ limit. One can then use a superconformal Ward identity to obtain the dynamical coefficient, $C_T$, that controls the two-point-function of the energy momentum tensor of the SCFT at large~$N$ 
\begin{equation}
C_T = \frac{32}{\pi^2} \frac{\partial^2F_{S^3_b}}{\partial b^2}|_{b=1} = \frac{32}{\pi^2} N^2 T\,.
\end{equation}

\item Another interesting choice for $\mathcal{M}_3$ is the manifold $S^1 \times \Sigma_{\mathfrak{g}}$ where $\Sigma_{\mathfrak{g}}$ is a smooth compact Riemann surface of genus $\mathfrak{g}$. The partition function of the SCFT on this background is referred to as the topologically twisted index, reflecting the fact that there is a particular background flux for the $\U(1)_R$ R-symmetry that implements Witten's topological twist on $\Sigma_{\mathfrak{g}}$. Employing the results above we conclude that the large $N$ topologically twisted index for these S-fold SCFTs should read
\begin{equation}
F_{S^1 \times \Sigma_{\mathfrak{g}}} = (1-\mathfrak{g})\frac{N^2 T}{2} \,.
\end{equation}
The 4d gravitational solutions dual to this index is the Euclidean Romans solution which admits a Lorentzian interpretation, when $\mathfrak{g}>1$, as the extremal supersymmetric Reissner-Nordstr\"om black hole in AdS$_4$, see \cite{Bobev:2021oku} for more details. The topologically twisted index serves as the counting function that accounts for the Bekenstein-Hawking entropy of this black hole.

\item The final example we consider is the supersymmetric limit of the Euclidean Kerr-Newman black hole. The 3d SCFT partition function in this case corresponds to the so-called superconformal index which counts particular supersymmetric operators in the SCFT. This index can also be computed via a path integral of the 3d theory on the manifold $S^1 \times_{\omega} S^2$ where $\omega$ is a real number associated with the angular momentum fugacity on $S^2$. Using the known regularized on-shell action for this gravitational background, see \cite{Bobev:2021oku} for a summary, we arrive at the following prediction for the large $N$ superconformal index of the family of 3d $\mathcal{N}=2$ S-fold SCFTs discussed in this work
\begin{equation}
F_{S^1 \times_{\omega} S^2} = \frac{(\omega+1)^2}{2\omega}\frac{N^2 T}{2} \,.
\end{equation}
This result for the index, after an appropriate choice of ensemble and a Legendre transformation, can also be used to obtain the Bekenstein-Hawking entropy of the supersymmetric Kerr-Newman black hole.

\end{itemize}

It will be very interesting to revisit the supersymmetric localization analysis in \cite{Assel:2018vtq} and understand whether the path integral of the S-fold SCFT on these and other Euclidean 3-manifolds can be computed at large $N$. This will open the door to exploring precision holography in this context and give access to the observables discussed above beyond the leading order in the large $N$ approximation.

\bigskip
\bigskip
\leftline{\bf Acknowledgements}
\smallskip
\noindent We are grateful to Chris Beem, Junho Hong, Krzysztof Pilch, Valentina Puletti, and Valentin Reys for fruitful discussions. The research of NB is supported by the FWO projects G003523N, G0E2723N, and G094523N, and Odysseus grant G0F9516N, as well as by the KU Leuven C1 grant ZKD1118 C16/16/005. FFG is supported by the Icelandic Research Fund under grant 228952-052 and is partially supported by grants from the University of Iceland Research Fund. JvM is supported by the ERC-CoG grant NP-QFT No. 864583 ``Non-perturbative dynamics of quantum fields: from new deconfined phases of matter to quantum black holes'' and by INFN Iniziativa Specifica ST\&FI.

\appendix

\section{The IIB background for $Y\rightarrow 0$ }
\label{App: Y to zero background}

To understand the properties of the conformal manifold of $\mathcal N=2$ S-folds we study the geometry in the limit $Y\rightarrow 0$. First we analyze the limit of the functions defined in \eqref{Eq: functions needed for the uplift} and find that the useful quantities that remain are
\begin{equation}
	\begin{aligned}
		(a_2^\pm)^{(0)} =& \sqrt{2}(\cos 2\xi_3 +(1\pm \sqrt{2}) (\cos 2\xi_3 - \sqrt{2} \cos 2\phi) \sin^2\theta)+ \mathcal O(Y)\,,\\
		(a_3^\pm)^{(0)} =& -4 + (2+\pm \sqrt{2} \cos 2\xi_3)\cos^2\theta \mp 2\cos 2\phi \sin^2\theta+ \mathcal O(Y) \,.
	\end{aligned}
\end{equation}
The ten-dimensional warp factor becomes
\begin{eqnarray}
	\Delta^{(0)} = -32 \cos^2\theta(\cos^2 2\xi_3 \cos^2\theta + \sqrt{2} \cos 2\xi_3 \cos 2\phi \sin^2\theta - 2) + \mathcal O(Y)\,.
\end{eqnarray}
For the metric on the internal space we find that
\begin{equation}
	\begin{aligned}
		(\dd s_5^2)^{(0)}=& \frac{64}{g^2 \Delta^{(0)}} \Big[ 2\cos^4\theta \Big( \tfrac14 \sin^2 2\xi_3 \dd\xi_2^2 + (1+\sin^2 2\xi_3)\dd\xi_3^2 \Big)  \\
		& + \sin^2\theta (2+\sqrt{2} \cos 2\xi_3 \cos 2\phi \cos^2\theta + 2 \sin^22\phi \sin^2\theta)\dd\phi^2\\
		& + \sqrt{2}\sin 2\xi_3 \sin2\phi \sin^22\theta \,\dd \xi_3 \dd\phi\\
		& +\sin 2\theta \Big( \cos^2\theta \sin 2\xi_3 (\cos 2\xi_3 - \sqrt{2} \cos 2\phi )\dd \xi_3 \\
		&- \sqrt{2}\sin2\phi (\sqrt{2} \cos 2\phi \sin^2\theta - \cos 2\xi_3 (1+\sin^2\theta))\dd \phi \Big) \dd\theta\\
		& + 2\cos^2\theta \Big(1-\tfrac{1}{\sqrt{2}} \cos 2\xi_3 \cos 2\phi  (3\sin^2\theta +1) \\
		&\hspace{2cm}+ (\cos^2 2\xi_3 + \cos^2 2\phi) \sin^2\theta\Big)\dd\theta^2 \Big] + \mathcal O(Y)\,.
	\end{aligned}
\end{equation}
The full ten-dimensional metric is then given by
\begin{equation}
	\dd s_{10}^2 = \frac{(\Delta^{(0)})^{1/4}}{g^2}\left[\left(2^{4/3} \dd \varphi^2\right)\frac{1}{Y^2} +\left( 2^{7/3} \dd s_{\text{AdS}_4}^2 +\frac{1}{2} (\dd s_5^2)^{(0)}\right) + \mathcal O(Y)\right]\,.
\end{equation}
The important point here is that the only place where the parameter $Y$ shows up is in front of the S-fold circle parameterized by $\varphi$. For the axion and dilaton we find that they also become independent of $Y$ in this limit:
\begin{equation}
	\begin{aligned}
		\rme^{\hat \Phi^{(0)}} =&\, \frac{4 \rme^{2\varphi}(2-\sqrt{2} \cos 2\xi_3 \cos^2\theta - 2 \cos^2 \phi \sin^2\theta)}{\sqrt{\Delta^{(0)}}} + \mathcal O(Y)\,,\\
		\hat C_0^{(0)} =&\, \frac{\rme^{-2\varphi} \sin2\theta_4 \sin^2\theta}{(2-\sqrt{2} \cos 2\xi_3 \cos^2\theta - 2 \cos^2 \phi \sin^2\theta)} + \mathcal O(Y)\,.
	\end{aligned}
\end{equation}
For the R-R and NS-NS forms we find
\begin{equation}
	\begin{aligned}
		\hat C_2^{(0)} =& \frac{16 \rme^{-\varphi}}{g^2}w\Big[ (\sqrt{2} + \cos 2\xi_3) \cos^2\theta \sin 2\xi_3 \dd \xi_2\wedge \dd \xi_3 \\ 
		&+ \frac12\cot\phi \Big( -(1-\sqrt{2}) \cos^2 \xi_3 ((a_3^-)^{(0)} - 4\sqrt{2} \sin^2\phi \sin^2\theta ) \dd(\xi_1 +  \xi_2)\\
		&\hspace{1cm} + (1+\sqrt{2}) \sin^2 \xi_3 ( (a_3^-)^{(0)}  + 4\sqrt{2} \sin^2\phi \sin^2\theta ) \dd(\xi_1 -  \xi_2) \Big) \wedge \dd\phi \\
		& + \frac{1}{\sin 2\theta} \Big( - (1-\sqrt{2})\cos^2 \xi_3 \cos^2\theta ((a_2^+)^{(0)} +2 ) \dd(\xi_1 +  \xi_2) \\
		&\hspace{1cm}+ (1+\sqrt{2})\cos^2\theta \sin^2\theta_3((a_2^-)^{(0)} +2 )\dd(\xi_1 -  \xi_2)  \Big) \Big] + \mathcal O(Y)\,,
	\end{aligned}
\end{equation}

\begin{equation}
	\begin{aligned}
		\hat B_2^{(0)} =& \frac{16 \rme^{\varphi}}{g^2}\partial_{\theta_4} w \Big[ -(\sqrt{2} - \cos 2\xi_3) \cos^2\theta \sin 2\xi_3 \dd\xi_2  \wedge \dd \xi_3 \\
		& + \frac12\tan\phi \Big( -(1+\sqrt{2}) \cos^2 \xi_3 ((a_3^+)^{(0)} + 4\sqrt{2} \cos^2 \phi \sin^2\theta) \dd(\xi_1 +  \xi_2) \\
		&\hspace{1cm} +(1-\sqrt{2}) \sin^2\xi_3 ((a_3^+)^{(0)} - 4\sqrt{2} \cos^2 \phi \sin^2\theta) \dd(\xi_1 -  \xi_2) \Big)\wedge \dd\phi \\
		&+ \frac12\cot\theta \Big(- (1+\sqrt{2}) \cos^2 \xi_3 (2-(a_2^-)^{(0)}) \dd(\xi_1 +  \xi_2) \\
		&+ (1-\sqrt{2}) \sin^2\theta_3 (2- (a_2^{+})^{(0)})\dd(\xi_1 -  \xi_2) \Big)\wedge\dd\theta \Big] + \mathcal O(Y)\,.
	\end{aligned}
\end{equation}
And finally, the four form is
\begin{equation}
	\begin{aligned}
		C_4^{(0)} =& \frac{16 \cos^2\theta \sin 2\xi_3 \sin 2\theta }{g^4} \Big[\sin 2\theta (  \sqrt{2} \cos 2\xi_3 \cos 2\phi \cos2\theta + 2\\
		 &\hspace{1cm} - 2 \cos^2 2\xi_3 \cos^2\theta)\dd \xi_3 \wedge \dd \phi - 2 \Big(  \sqrt{2}  \cos 2\phi \cos 2\xi_3 \sin^2\theta \dd \phi \\
		&\hspace{1cm}  - \cos 2\xi_3 \cos^2\theta (\sqrt{2} \sin2\phi \dd \xi_3 + 2 \sin 2\xi_3 \dd\phi)  \Big) \wedge \dd\theta \Big] \wedge \dd\xi_1 \wedge\dd\xi_2 + \mathcal O(Y)\,.
	\end{aligned}
\end{equation}
%

\section{Four parameter family of AdS$_4$ solutions}\label{App: non-susy sols}
In this appendix we further extend the family of supergravity solutions, dual to the conformal manifold, to include a fourth independent parameter. As shown in Section~\ref{Sec: 5d}, the three parameter family of backgrounds is already rather involved. It is outside the scope of this paper to provide a type IIB supergravity description of the four-dimensional family of solutions, instead we will utilize four-dimensional gauged supergravity to construct the full four-dimensional manifold of solutions. In particular, we will use $[\SO(6)\times \SO(1,1)] \ltimes \mathbf R^{12}$ gauged, $\mathcal N=8$ supergravity in four dimensions. The gauging is specified by an embedding tensor ${X_{\bb M\bb N}}^{\bb R}$ \cite{deWit:2007kvg}  where $\bb M=1,\cdots,56$. The embedding tensor takes the explicit form
\begin{equation}\label{Eq: 4d embedding tensor}
\begin{aligned}
	&X_{[AB][CD]}^{\phantom{[AB][CD]}[EF]} = - X_{[AB]\phantom{[EF]}[CD]}^{\phantom{[AB]}[EF]} = - 8 \delta_{[A}^{[E}\theta_{B][C}\delta_{D]}^{F]}\,,\\
	&X^{[AB]\phantom{[EF]}[EF]}_{\phantom{[AB]}[CD]} = -X^{[AB][EF]}_{\phantom{[AB][EF]}[CD]} = - 8 \delta_{[C}^{[A}\xi^{B][E}\delta_{D]}^{F]}\,,
\end{aligned}
\end{equation}
where $A,B,\ldots$ are $\mathfrak{sl}(8)$ indices running from $1$ to $8$. For the $\SO(6)\times \SO(1,1) \ltimes \mathbf R^{12}$ gauging, the $8\times 8$ matrices $\theta$ and $\xi$ take the form
\begin{equation}\label{thetaandxi}
	\theta = g \, \text{diag}(1,1,1,1,1,1,0,0)\,,\quad \xi = g \, \text{diag}(0,0,0,0,0,0,-1,1)\,,
\end{equation}
with $g$ the gauge coupling constants. The scalar sector of this theory consists of 70 scalars, describing the coset $E_{7(7)}/\SU(8)$. We use the real $\SL(8)$  basis of $\Ess$ as in \cite{Cremmer:1979up,Guarino:2015qaa,Bobev:2020qev,Guarino:2021hrc}. For a given $\ssl(8)$ matrix ${\Lambda_A}^B$ and four-form $\Sigma_{ABCD}$, the corresponding $\ess$ generator can be constructed as
\begin{equation}\label{Eq: generators 4d sugra}
	\mathfrak X_{\bb M}^{\phantom{\bb M}\bb N}(\Lambda,\Sigma) = \left(\begin{matrix}{}
  2 \Lambda_{[A}^{\phantom{[A}[C} \delta_{B]}^{D]} & \Sigma_{[AB][CD]}\\
  \Sigma^{[AB][CD]} & -2 \Lambda_{[C}^{\phantom{[C}[A} \delta_{D]}^{B]}
\end{matrix}\right)\,.
\end{equation}
Using these generators, a generic element ${\cal V}$ of the coset space can be parametrized by a product of exponentials.
The metric on the coset manifold is then given by 
\begin{equation}
\mathcal M = \mathcal V \cdot \mathcal V^T\,.
\end{equation}
Using the technology in \cite{deWit:2007kvg} one can readily compute the Lagrangian associated to this truncation. In particular, the potential on the scalar manifold takes the form
\begin{equation}
	V = \frac{1}{672} \mathcal M^{\bb M\bb P} X_{\bb M\bb N}^{\phantom{\bb M\bb N}\bb R} X_{\bb P\bb Q}^{\phantom{\bb P\bb Q}\bb S} (\mathcal M^{\bb N\bb Q} \mathcal M_{\bb R\bb S} + 7 \delta_{\bb S}^{\bb N} \delta_{\bb R}^{\bb Q} )\,,
\end{equation}
and the family of vacua we are after are extrema of this potential, with value
\begin{equation}
	V = - 3 g^2\,.
\end{equation}
A two-dimensional parameter space of such vacua was found in \cite{Bobev:2021yya} by using a $\mathbf Z_2^3$ truncation on the scalar coset. We will use a slightly different construction here which allows us to extend this solution to a four-dimensional family of solutions. We start by noting that the form of the embedding tensor specified in \eqref{thetaandxi} implies that the $\SO(6)$ gauge group is embedded in $\SL(8)$ in the upper left block
\be
\SL(8)\supset\begin{bmatrix} \SO(6) &\\&\SO(1,1)\end{bmatrix}\,.
\ee
We look for $\U(1)\times \U(1)$ invariant critical points of the potential, where the two $\U(1)$'s are embedded in $\SO(6)$ as specified by the following generators
\be\label{eq:defg1g2}
g_1 = \f12 \begin{bmatrix}
0 &0 &0&1&0&0\\
0 &0 &1&0&0&0\\
0 &-1&0&0&0&0\\
-1&0 &0&0&0&0\\
0 &0 &0&0&0&0\\
0 &0 &0&0&0&0\end{bmatrix}\,,\qquad g_2 = \f12 \begin{bmatrix}
0 &0 &1&0&0&0\\
0 &0 &0&1&0&0\\
-1&0 &0&0&0&0\\
0 &-1&0&0&0&0\\
0 &0 &0&0&0&0\\
0 &0 &0&0&0&0\end{bmatrix}\,.
\ee
As discussed in \cite{Bobev:2021yya} the $\U(1)\times\U(1)$ truncation of the four-dimensional theory contains 18 scalar fields parametrizing the coset
\be\label{u1u1invariantmfd}
{\cal M} = \f{\SL(2,\R)}{\U(1)}\times \f{\SO(4,4)}{\SO(4)\times\SO(4)}\,.
\ee
The four-dimensional solution we are after is contained in this scalar manifold, but is difficult to construct analytically. Fortunately there is a simplification we can employ. Two out of the four parameters come `for free' using the result of \cite{Guarino:2021hrc} because we impose $\U(1)\times\U(1)$ invariance. The results of \cite{Guarino:2021hrc} state that for any critical point ${\cal V}_0$ which is invariant under some $H\subset \SO(6)$ there are $\text{rk}(H)$ flat deformation of the seed solution ${\cal V}_0$ that leave the potential invariant. The flat deformation can be constructed as follows. Let $g$ be a element of the Cartan subalgebra of $H$ and embed that into $\SL(8)$ as a two-form, then let
\be
\Sigma^{(g)} = \dd x^7\wedge \dd x^8\wedge g \,,
\ee
here we use $x^i$ as coordinates on the auxiliary $\R^8$ where the $\SL(8)$ forms live. The $\Ess$ generator $\mathfrak{h} = \mathfrak X(0,\Sigma^{(g)})$ is a flat direction in the sense that ${\cal V} = \e^{\chi \mathfrak h} \cdot {\cal V}_0$ is a critical point with the same value of the potential for any $\chi$. For our $\U(1)^2$ symmetry there will be two such flat directions we can utilize to simplify our construction. In particular we can impose additional discrete symmetry and construct a two-dimensional subspace of solutions, and then generate the remaining two directions as described. We use the $\Z_2$ symmetry embedded in $\SO(6)$ as follows
\be
S = \begin{bmatrix}
-1&0 &0&0&0&0\\
0 &-1&0&0&0&0\\
0 &0 &1&0&0&0\\
0 &0 &0&1&0&0\\
0 &0 &0&0&-1&0\\
0 &0 &0&0&0&-1\end{bmatrix}\,.
\ee
Truncating further to the invariant subspace of $S$ the scalar manifold simplifies
\be\label{u1u1z2mfd}
 \f{\SO(4,4)}{\SO(4)\times\SO(4)} \to  \f{\SO(3,3)}{\SO(3)\times\SO(3)}\times \O(1,1)\,.
\ee
Notice that the $\SL(2,\R)/\U(1)$ factor in \eqref{u1u1invariantmfd} is left invariant by the discrete symmetry. However we find that the condition for finding a critical point sets all parameters of that factor to zero and so we only have to specify the location within the scalar manifold on the right-hand-side of \eqref{u1u1z2mfd}. 

Even though this manifold is ten-dimensional, the family of critical points can be specified in terms of only six generators. These are
\begin{equation}
\begin{aligned}
&\mathfrak t_1 = \mathfrak X\big(\text{diag}(0,0,0,0,0,1,0,-1),0\big)\,,\qquad \mathfrak t_2 = \mathfrak X\big(\text{diag}(0,0,0,0,0,-1,1,0),0\big)\,,\\
&\mathfrak r_1 = \mathfrak X\big(0, g_1\wedge \dd x^6\wedge \dd x^7 \big)^T\,,\qquad \mathfrak r_2 = \mathfrak X\big(0, g_2\wedge \dd x^6\wedge \dd x^7 \big)^T\,,\\
&\mathfrak r_3 = \mathfrak X\big(0, g_1\wedge \dd x^5\wedge \dd x^7 \big)^T\,,\qquad \mathfrak r_4 = \mathfrak X\big(0, g_2\wedge \dd x^5\wedge \dd x^7 \big)^T\,.
	\end{aligned}
\end{equation}
To simplify the expression for the scalar vielbein we define two field-dependent generators:
\be
\begin{split}
\mathfrak A &= -\f{\sqrt{(1+x)(1-z)}\,\mathfrak r_1 + \sqrt{(1-x)(1+z)}\,\mathfrak r_2}{\sqrt{4+2x+2z}}\,,\\
\mathfrak B &= \f{-\sqrt{1-z^2}\,\mathfrak r_3 + \sqrt{1-x^2}\,\mathfrak r_4}{\sqrt{(2+x+z)(4+x+z)}} \,.
\end{split}
\ee
The two-parameter seed solution can now be specified by the 56-bein
\begin{equation}
\begin{aligned}
\mathcal V_0 =&\, \rme^{-\log \sqrt{2} \mathfrak t_1} \cdot \rme^{\f12\log\frac{4+x+z}{4} \mathfrak t_2} \cdot \rme^{\f{2}{-2+x+z}[\mathfrak A,\mathfrak B]}\cdot \rme^{\mathfrak A} \cdot \rme^{\mathfrak B} \,.
\end{aligned}
\end{equation}
We have verified that this is a critical point by evaluating the gradients of the potential in all 70 directions of the maximal supergravity theory. Another way to verify criticality would be to construct the potential of the $\U(1)\times\U(1)\times\Z_2$ truncation and demonstrate that the potential is extremized by the above solution.

Before extending this solution to a four-parameter family let us point out a few of its features. First note that interchanging $x$ and $z$ leads to the same solution and is therefore a symmetry. The special line of solutions $x=z$ has enhanced $\SU(2)\times \U(1)$ continuous symmetry which gets enhanced to $\SO(5)\times \SO(1,1)$ at $x=z=1$ and it gets enhanced to ${\cal N}=4$ supersymmetry with $\SO(4)$ R-symmetry at $x=z=-1$. Except for this one point the line $x=z$ is non-supersymmetric. A one-dimensional ${\cal N}=2$ supersymmetric family first identified in \cite{Bobev:2021yya} can be obtained by setting
\be
z=-1\,,\qquad  \text{or}\qquad x=-1\,.
\ee
For $z=-1$ the ${\cal N}=2$ R-symmetry is generated by $g_1$ whereas for $x=-1$ it is $g_2$ that generates the R-symmetry. Along the ${\cal N}=2$ supersymmetric lines we have two enhancement points. One is the ${\cal N}=4$ supersymmetric point at $x=z=-1$, the other is located at $x=-1$ and $z=1$ or $x=1$ and $z=-1$. In Section \ref{Sec: 5d} we uplifted this supersymmetric family in addition to the two simple flat directions discussed below. In order to make contact with section \ref{Sec: 5d} the parameters used in this appendix are related to the parameters used there as follows
\be
x=2Y^2-3\,,\qquad z=-1\,.
\ee

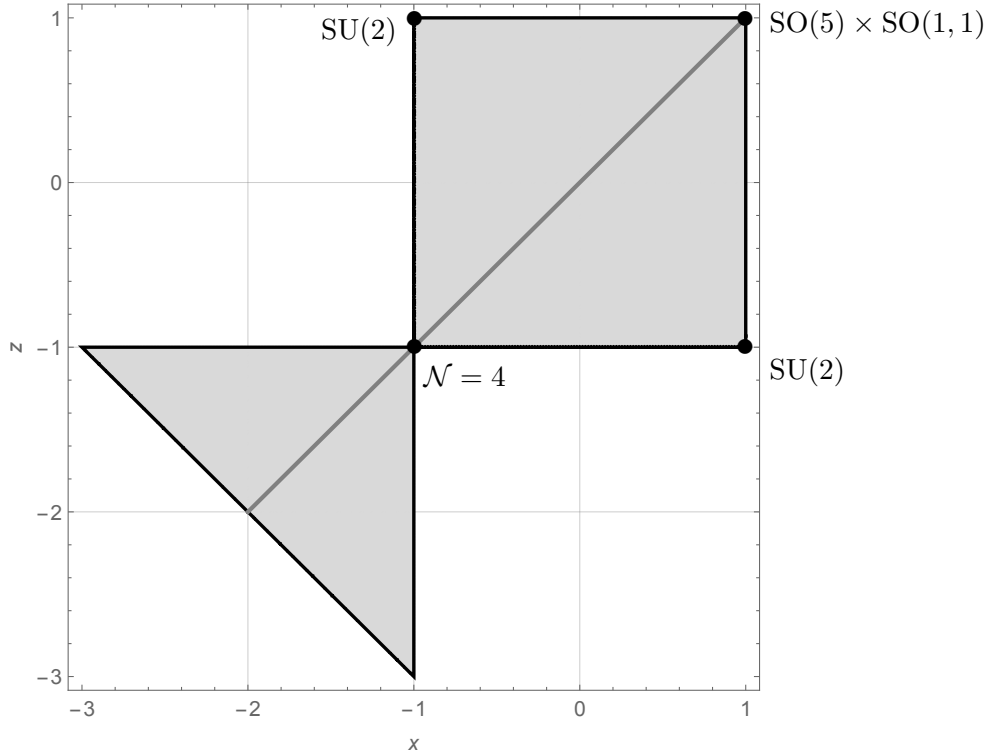
\begin{figure}[ht]
\centering
\begin{overpic}[width=10cm]{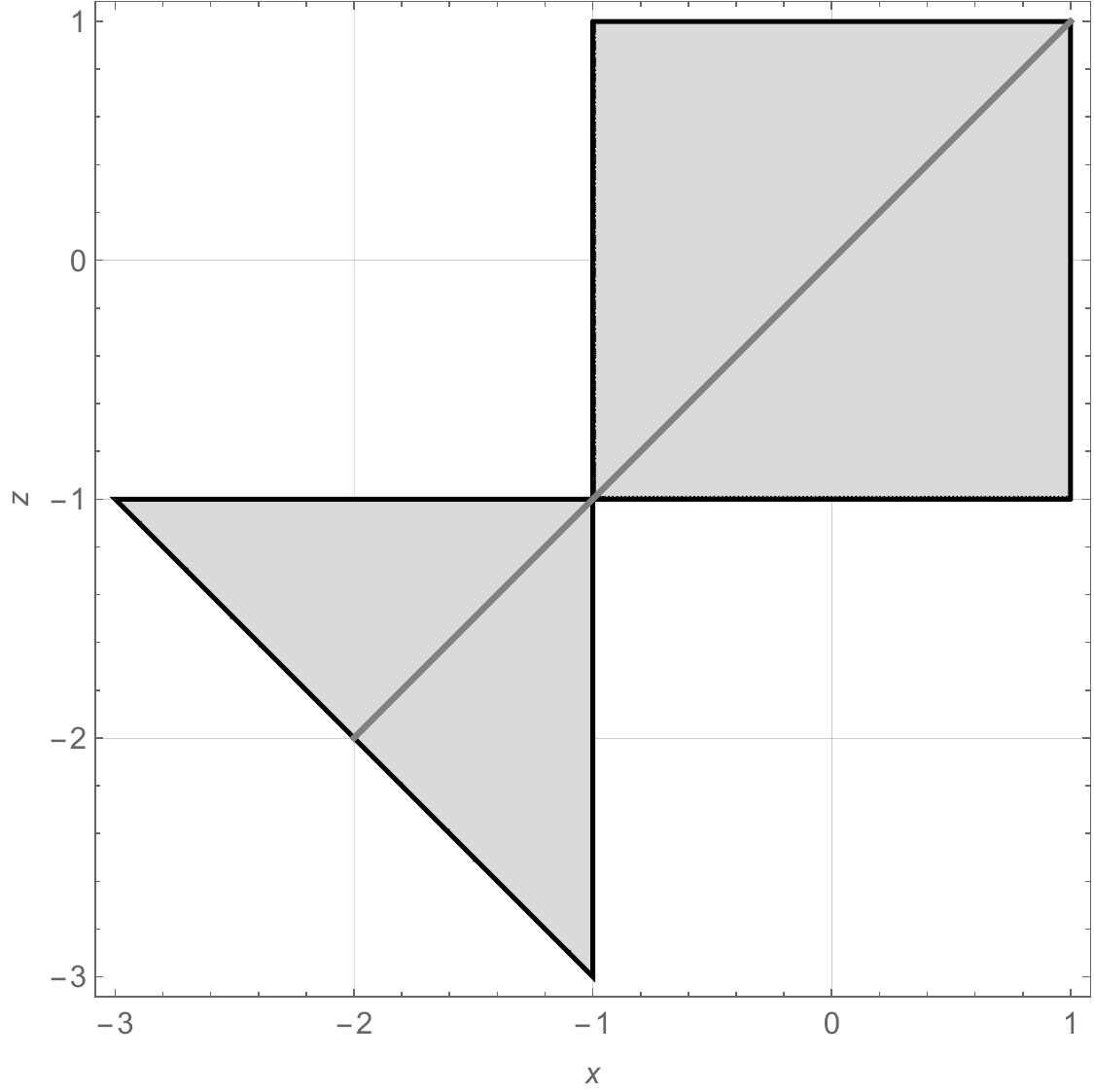}
\put(53,53){\Large$\bullet$}
\put(55.5,49){${\mathcal N}=4$}
\put(96.5,96.5){\Large$\bullet$}
\put(101,96){$\text{SO}(5)\times\text{SO}(1,1)$}
\put(53,96.5){\Large$\bullet$}
\put(42,95){$\text{SU}(2)$}
\put(96.5,53){\Large$\bullet$}
\put(101,50){$\text{SU}(2)$}
\end{overpic}
\caption{\label{CMregions}The allowed region for the two parameters $x$ and $z$. The solution is symmetric under the interchange of $x$ and $z$ and the dark-gray diagonal line is the axis of symmetry. Along this axis the continuous symmetry is enhanced from $\text{U}(1)^2$ to $\text{SU}(2)\times\text{U}(1)$. The ${\mathcal N}=2$ supersymmetric lines are at $z=-1$ or $x=-1$ -- both represent the same solution. Other symmetry enhanced points are marked on the figure.}
\end{figure}

In Figure~\ref{CMregions} we display the allowed range of parameters $x$, $z$ and also indicate the symmetry enhancement points.

In order to extend the above solution to a full four-parameter family, we use the procedure outlined above and write
\be\label{full4Dsol}
{\cal V} = \e^{ \chi_1 \mathfrak h_1}\cdot \e^{ \chi_2 \mathfrak h_2}\cdot{\cal V}_0\,,
\ee 
where the generators $\mathfrak h_{1,2}$ are defined by
\be
\mathfrak h_1 = \mathfrak X\big(0,g_1\wedge \dd x^7\wedge \dd x^8\big)\,,\qquad
\mathfrak h_2 = \mathfrak X\big(0,g_2\wedge \dd x^7\wedge \dd x^8\big)\,.
\ee
Even though $x$ and $z$ may be chosen such that ${\cal V}_0$ exhibits enhanced symmetry, the deformation in \eqref{full4Dsol} generically brakes the bosonic symmetry to $\U(1)\times\U(1)$. In order to maintain supersymmetry the parameter $\chi_i$ should be set to zero for the $\mathfrak h_i$ that corresponds to the R-symmetry. Note that these $\chi_i$ parameters can be directly identified with those introduced in ten dimensions in Section~\ref{Sec: Two parameter family of novel type IIB vacua}.

\bibliography{N=2CMin10d}
\bibliographystyle{utphys}

\end{document}